\documentclass[useAMS,usenatbib]{mn2e}

\usepackage{natbib}
\usepackage{graphicx}

\newcommand{\beq}{\begin{equation}}
\newcommand{\eeq}{\end{equation}}
\newcommand{\beqa}{\begin{eqnarray}}
\newcommand{\eeqa}{\end{eqnarray}}

\title[Magnetic swirls in a solar flux tube]
{Magnetic swirls and associated fast magnetoacoustic kink waves in a solar chromospheric flux tube}

\author[K. Murawski et al.]
  {K.~Murawski$^1$,
   P.~Kayshap$^1$,
   A.~K.~Srivastava$^{3}$,
   D.~J.~Pascoe$^2$,
   P.~Jel\'\i nek$^{4}$,
   B.~Ku\'zma$^1$ \and and
   V.~Fedun$^{5}$\\   
  $^1$Maria Curie-Sk{\l}odowska University, Institute of Physics, Group of Astrophysics, Radziszewskiego 10, PL--20 031 Lublin, Poland \\
  $^2$Centre for Fusion, Space and Astrophysics, Department of Physics, University of Warwick, CV4 7AL, UK \\
  $^3$Department of Physics, Indian Institute of Technology (Banaras Hindu University), Varanasi-221005, India \\
  $^4$University of South Bohemia, Faculty of Science, Institute of Physics and Biophysics, Brani\v sovsk\'a 1760, CZ -- 370 05 \\ 
  \v{C}esk\'e Bud\v{e}jovice, Czech Republic \\
  $^5$Space Systems Laboratory, Department of Automatic Control and Systems Engineering, University of Sheffield, Sheffield, S1 3JD, UK
}

\date{Released 2017 Xxxxx XX}

\pagerange{\pageref{firstpage}--\pageref{lastpage}} \pubyear{2017}

\def\LaTeX{L\kern-.36em\raise.3ex\hbox{a}\kern-.15em
    T\kern-.1667em\lower.7ex\hbox{E}\kern-.125emX}

\begin{document}

\label{firstpage}

\maketitle

\begin{abstract}
We perform numerical simulations of impulsively generated magnetic swirls in an isolated flux tube which is rooted in the solar photosphere. 
These swirls are triggered by an initial pulse in a horizontal component of the velocity.
The initial pulse is launched either: (a) centrally, within the localized magnetic flux tube; or (b) off-central, in the ambient medium.
The evolution and dynamics of the flux tube is described by three-dimensional, ideal magnetohydrodynamic equations.  
These equations are numerically solved to reveal that in case (a) dipole-like swirls associated with the fast magnetoacoustic kink and $m=1$ Alfv\'en waves are generated.
In case (b), the fast magnetoacoustic kink and $m=0$ Alfv\'en modes are excited. 
In both these cases, the excited fast magnetoacoustic kink and Alfv\'en waves consist of similar flow pattern and 
magnetic shells are also generated with clockwise and counter-clockwise rotating plasma within them, which can be the proxy of dipole-shaped chromospheric swirls.
The complex dynamics of vortices and wave perturbations reveals 
the channelling of sufficient amount of energy to fulfill energy losses in the chromosphere ($\sim$ 10$^{4}$ W m$^{-1}$) and in the
corona ($\sim$ 10$^{2}$ W m$^{-1}$).
Some of these numerical findings are reminiscent of signatures in recent observational data.
\end{abstract}


\section{Introduction}
The solar atmosphere consists of three main layers with the chromosphere being sandwiched between the photosphere and the solar corona.
This magnetically structured layer plays an important role in transporting energy and mass from the photosphere to the corona.
The localized magnetic fields and their complex structuring support the formation of various plasma jets (e.g. spicules, anemone jets, swirls), 
and also result in excitation, trapping, and conversion of various magnetohydrodynamic (MHD) waves 
\citep[e.g.][and references therein]{Bonet2008, 2004Natur.430..536D, 2014Sci...346D.315D, 1995SoPh..162..233H, 2007Sci...318.1591S, 2008A&A...481L..95S, 2007A&A...461L...1V, 2013MNRAS.434.2347J, 2014RAA....14..805P, 2014Sci...346A.315T, 2015A&A...581A.131J,Srivastava2017}.

The chromospheric mass supply and heating depend upon many localized and dynamical plasma processes 
which include small-scale jets, swirls, and network jets \citep[e.g.][and references therein]{Wed2009,Murawski2011,Wed2012,2014Sci...346A.315T,Rao2017,Giagkiozis2017}. 
Chromospheric swirls were considered as potential heating and mass supply candidates by \citet{Wed2012} and their classification and various morphologies (e.g. ring- and spiral-shaped swirls) was later established in the solar chromosphere by \citet{Wed2013}.
These swirls were found to be driven by solar convective motions \citep[e.g.][]{2010ApJ...723L.180S, 2011A&A...526A...5S, Wed2012}.
\citet{2014ApJ...788....8M} proposed a new physical mechanism for the generation of such chromospheric swirls. They reported that these ejecta are associated with the propagation of torsional Alfv\'en waves which result in concentric magnetic shells and vertical plasma flows in the non-linear regime.

These swirls and vortices are mostly associated with the quiet Sun (QS), which is filled by granules at the photosphere and in magnetic networks.
Central regions of granules show upflows while their edges (i.e. near the intergranular lanes) exhibit downflows, and the intermediate region experiences horizontal flows \citep[e.g.][]{Svanda2006,Zhao2007,Verma2011,Rou2012,Rou2013,Svanda2013}.
\citet{Rie2001} found that horizontal plasma flows and granular/supergranular motions are highly correlated at scales greater than 2.5~Mm.
Eddies (i.e. vortex flows) in the intergranular lanes in the QS were reported by \cite{Brandt1988}, which form due to convective flows in the presence of magnetic flux tubes (i.e. magnetic bright points; MBPs).
It is well-established that these vortex-like motions form due to the granular/supergranular horizontal flows \citep[e.g.][and references therein]{Wang1995,Duvall2000,Gizon2003, Bonet2008,Attie2009}.
\citet{Innes2009} detected vortices in the inflow regions (i.e. intergranular lanes) by calculating horizontal plasma flows, which are plasma motions from the centres of granules towards their edges.
Similarly, \citet{Attie2009} detected vortex flow at the supergranular junctions due to the horizontal plasma flow.

Interaction between vortex flow and magnetic flux tubes may lead to the formation of chromospheric swirls as well as magnetic tornadoes \citep[e.g.][]{Wed2009,Wed2012,Wed2013}.
The formation of vortex flows and their role in triggering various kind of MHD waves are fundamental questions of solar physics.
However, despite the availability of high resolution observations in addition to advanced numerical simulations, the formation of these vortex flows and associated wave dynamics is not well understood up to now.
Convective flow is the most widely used mechanism to account for the vortex flow in the solar photosphere \citep[e.g.][]{Brandt1988,Bonet2008,Wed2009,Wed2012,Wed2013,Kitiashvili2013}. 
Recently, \citet{Murawski2015} reported the formation of vortex flows which can be triggered by horizontal velocity flows. Buffeting of a flux tube (i.e. interaction between a horizontal velocity pulse, mimicking the horizontal flow, and an isolated magnetic flux tube, imitating an MBP in intergranular lanes) can lead to the formation of unique dipole-shaped swirls alongside the natural generation of kink and Alfv{\'e}n waves in the simulations.
\begin{figure}
\resizebox{\hsize}{!}{\includegraphics{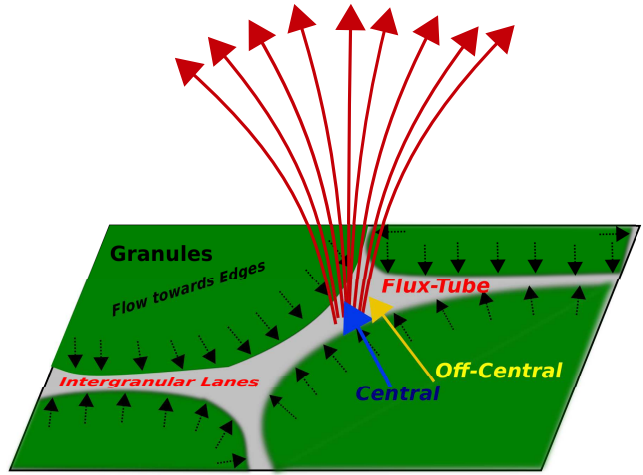}}
\caption{Schematic of the general structure of granules (green) and intergranular lanes (gray)
in a quiet Sun region of the solar photosphere.
Black arrows show the plasma flows towards the edges of the granules. 
Red lines indicate a magnetic flux tube, which is located in the intergranular lane.
To understand the formation of vortex motions we consider the two cases of buffeting by horizontal flows in the central part of the flux tube (blue arrow), and at one edge of the flux tube (yellow arrow).} 
\label{fig:cartoon}
\end{figure}

In this paper we investigate how changes in the horizontal flows (i.e. variations in strength, alterations of the interaction region) can affect the behaviour of vortex motions.
To address this question, we consider the physical implications of the generation of double 
quadrupole chromospheric swirls
in an isolated solar flux tube and their association with kink and Alfv\'en waves. 
This type of motion was recently detected by \citet{Giagkiozis2017}. 
We focus on two different locations for the horizontal velocity pulse with respect to the magnetic flux tube, as shown in Fig.~\ref{fig:cartoon}
which depicts a typical scenario in the QS solar photosphere.
Granules and intergranular lanes are represented by green and gray color patches, respectively.
Plasma flows from the central area of granules towards their edges (black arrows).
Red lines illustrate the isolated magnetic flux tube in an intergranular lane.
In the first case, we simulate the interaction between the flux tube and a centrally launched horizontal pulse at the top of the photosphere (blue arrow).
In the second case, we use an off-center horizontal velocity pulse (yellow arrow).
This paper is arranged as follows;
we present our three-dimensional (3D) numerical model of the solar atmosphere in Sect.~\ref{sect:model} and describe the plasma dynamics resulting from the applied pulses in Sect.~\ref{sect:results}.
A summary and further discussion are presented in Sect.~\ref{sect:summary}.
\section{Numerical model}\label{sect:model}
We consider a gravitationally-stratified solar atmosphere with an isolated magnetic flux tube that is described by the set of ideal, adiabatic, three-dimensional magnetohydrodynamic (MHD) equations in the Cartesian coordinate system.
The set of governing equations are given as

\beqa
\label{eq:MHD_rho}
{{\partial \varrho}\over {\partial t}}+\nabla \cdot (\varrho{\bf V})=0\, ,
\\
\label{eq:MHD_V}
\varrho{{\partial {\bf V}}\over {\partial t}}+ \varrho\left ({\bf V}\cdot \nabla\right ){\bf V} =
-\nabla p+ \frac{1}{\mu_{0}}(\nabla\times{\bf B})\times{\bf B} +\varrho{\bf g}
\, ,
\\
\label{eq:MHD_B}
{{\partial {\bf B}}\over {\partial t}}= \nabla \times ({\bf V}\times{\bf B})\, ,
\hspace{3mm}
\nabla\cdot{\bf B} = 0\, ,
\\
\label{eq:MHD_p}
{\partial p\over \partial t} + \nabla\cdot (p{\bf V}) = (1-\gamma)p \nabla \cdot {\bf V}\, ,
\hspace{3mm}
p = \frac{k_{\rm B}}{m} \varrho T\, .
\eeqa
Here $\varrho$ denotes the mass density, while ${\bf V}=(V_{\rm x},V_{\rm y},V_{\rm z})$ is the velocity vector, and ${\bf B}=(B_{\rm x},B_{\rm y},B_{\rm z})$ is the magnetic field.
The state of the plasma is also described by the gas pressure $p$ and temperature $T$, while
$\gamma=1.4$ is the adiabatic index,
${\bf g}=(0,-g_{\sun},0)$ as the gravitational acceleration with magnitude $g_{\sun}=274$~ms$^{-2}$,
$\mu_0 = 1.26 \times 10^{-6}$~Hm$^{-1}$ is the magnetic permeability of free space,
and $k_{\rm B}$ is the Boltzmann constant.
The symbol $m$ represents the mean particle mass which is specified as the product of the mean molecular weight of $1.24$ (Oskar Steiner; private communication) and the proton mass.
\subsection{Initial equilibrium}
\begin{figure}
\resizebox{\hsize}{!}{\includegraphics{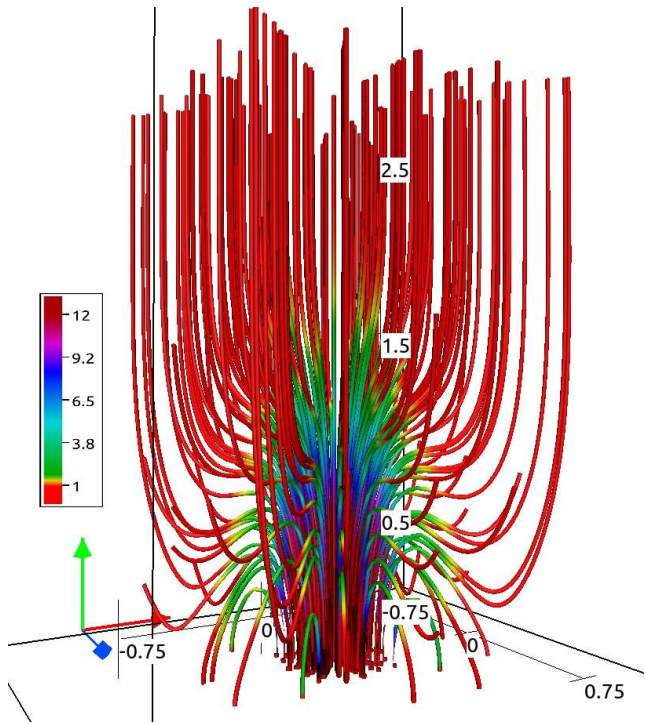}}
\caption{Equilibrium magnetic field lines.
Red, green, and blue arrows correspond to the $x$-, $y$-, and $z$-axes, respectively.
The size of the box shown is $(-0.75,0.75) \times (0,2.5) \times (-0.75,0.75)$~Mm and represents only part of the simulated region.
The color map corresponds to the magnitude of the magnetic field which is expressed in units of $B_{\rm y0} \approx 11.4$~G.}
\label{fig:initial_B}
\end{figure}
We consider an axisymmetric magnetic flux tube which is described in detail by \citet{2015A&A...577A.126M} who implemented the analytical theory developed by \citet{2010ARep...54...86S} and \citet{2015AstL...41..211S}.
The tube is initially non-twisted and the non-force-free 
magnetic field is represented by the azimuthal component of the magnetic flux function ($A\,{\bf\hat \theta}$) as
\beq\label{eq:equil_B}
{\bf B_{\rm }} = \nabla\times (A\,{\bf\hat \theta})\, ,
\eeq
where ${\bf\hat \theta}$ is a unit vector in the azimuthal direction.
In the axisymmetric case, we consider $\partial/\partial\theta = 0$. 
After this consideration, it follows from Eq.~(\ref{eq:equil_B}) that 
\beq\label{eq:B_com_A}
B_{\rm r} = -\frac{\partial A}{\partial y}\, ,\hspace{3mm} 
B_{\rm \theta} = 0\, , \hspace{3mm} 
B_{\rm y} = \frac{1}{r} \frac{\partial (rA)}{\partial r}\, ,
\eeq
where $r=\sqrt{x^2+z^2}$ is the radial coordinate in the horizontal $xz$-plane.
For our flux tube, we specify $A(r,y)$ as
\beqa\label{eq:A-Psi-flux tube}
A(r,y) = B_{\rm 0} \exp{\left( -k_{\rm y}^2 y^2 \right)} \frac{r}{1+k_{\rm r}^4r^4} + \frac{1}{2}B_{\rm y0} r \, ,
\eeqa
where $B_{\rm y0}$ is the magnitude of the ambient, vertical magnetic field while $k_{\rm r}$ and $k_{\rm y}$ are the inverse length scales along the radial and vertical directions, respectively.
We choose and hold fixed $k_{\rm r}=k_{\rm y}=4$~Mm$^{-1}$ and the magnitude of 
the magnetic field $B_{\rm 0}$ such that
the magnetic field within the flux tube, at $r=0$ and $y=0$, is $\approx 137$~G, and $B_{\rm y0} \approx 11.4$~G. 
The magnetic field lines corresponding to an isolated, non-twisted, axisymmetric magnetic flux tube are displayed in Fig.~\ref{fig:initial_B}.

The equilibrium mass density and plasma gas pressure 
are specified in Appendix and they are 
estimated using the temperature profile given by \cite{2008ApJS..175..229A}.
This profile is smoothly extended above the transition region and into the corona (see Murawski et al. 2015b for details).
\subsection{Perturbations to the flux tube}
Based on observations, it is natural to assume that the flux tube is continuously buffeted by a number of perturbations. 
In our study we consider a simple case in which the perturbation to the magnetic field is modelled by a Gaussian pulse in the horizontal component of the velocity, $V_{\rm x}$, 
\beqa\label{eq:perturb}
V_{\rm x}(x,y,z,t=0) = 
A_{\rm v}\, \exp {\left[-\frac{r_v^2}{w^2}\right]}\, ,
\eeqa
where $r_v^2=(x-x_{\rm 0})^2+(y-y_{\rm 0})^2+(z-z_{\rm 0})^2$,
$A_{\rm v}$ is the amplitude of the pulse, and
$(x_{\rm 0},y_{\rm 0},z_{\rm 0})$ and $w$ are its initial position and width, respectively.
We set and fix $A_{\rm v}=2$~km~s$^{-1}$, $w=0.125$~Mm, and $y_{\rm 0}=0.5$~Mm.
The value of $y_{\rm 0}$ corresponds to the top of the photosphere in our model solar atmosphere.
The impulsive generation of fast MHD waves in a 2D magnetic funnel was modelled by \cite{2013A&A...560A..97P,Pascoe2014}.
Earlier investigations, demonstrate that a pulse in velocity triggers waves and vortices \citep[e.g.][]{Murawski2015}.
The impulsive perturbation given by Eq.~(\ref{eq:perturb}) excites MHD-gravity waves propagating in the solar atmosphere.
\section{Results}\label{sect:results}
Equations (\ref{eq:MHD_rho}){--}(\ref{eq:MHD_p}) are solved numerically, using the \textsc{FLASH} code (Lee and Deane 2009; Lee 2013) 
to investigate the excitation and the behaviour of double swirls in an isolated magnetic flux tube. 
For all our simulations we use a numerical domain of size 
$(-1.25,1.25) \times (-1.25,1.25)$~Mm in horizontal directions and $(0, 20)$~Mm in vertical direction. 
We impose the chosen boundary conditions by fixing in time all the plasma parameters to their equilibrium values. 
The coarse numerical grid used near the top of the domain acts as a sponge, diffusing incoming waves.
The finest grids are used below the altitude $y=4$~Mm, where the effective finest spatial resolution is $\approx 0.0195$~Mm.

We consider two particular cases of Eq.~(\ref{eq:perturb}) in this paper;
a centrally-launched initial pulse which corresponds to $x_{\rm 0}=z_{\rm 0}=0$~Mm, generating double swirls and associated waves, and
an off-center initial pulse with $x_{\rm 0}=0$~Mm and $z_{\rm 0}=0.3$~Mm, generating swirls and associated waves.
\subsection{Perturbation by a centrally-launched pulse}
\begin{figure*}
\centering
\includegraphics[width=8.5cm]{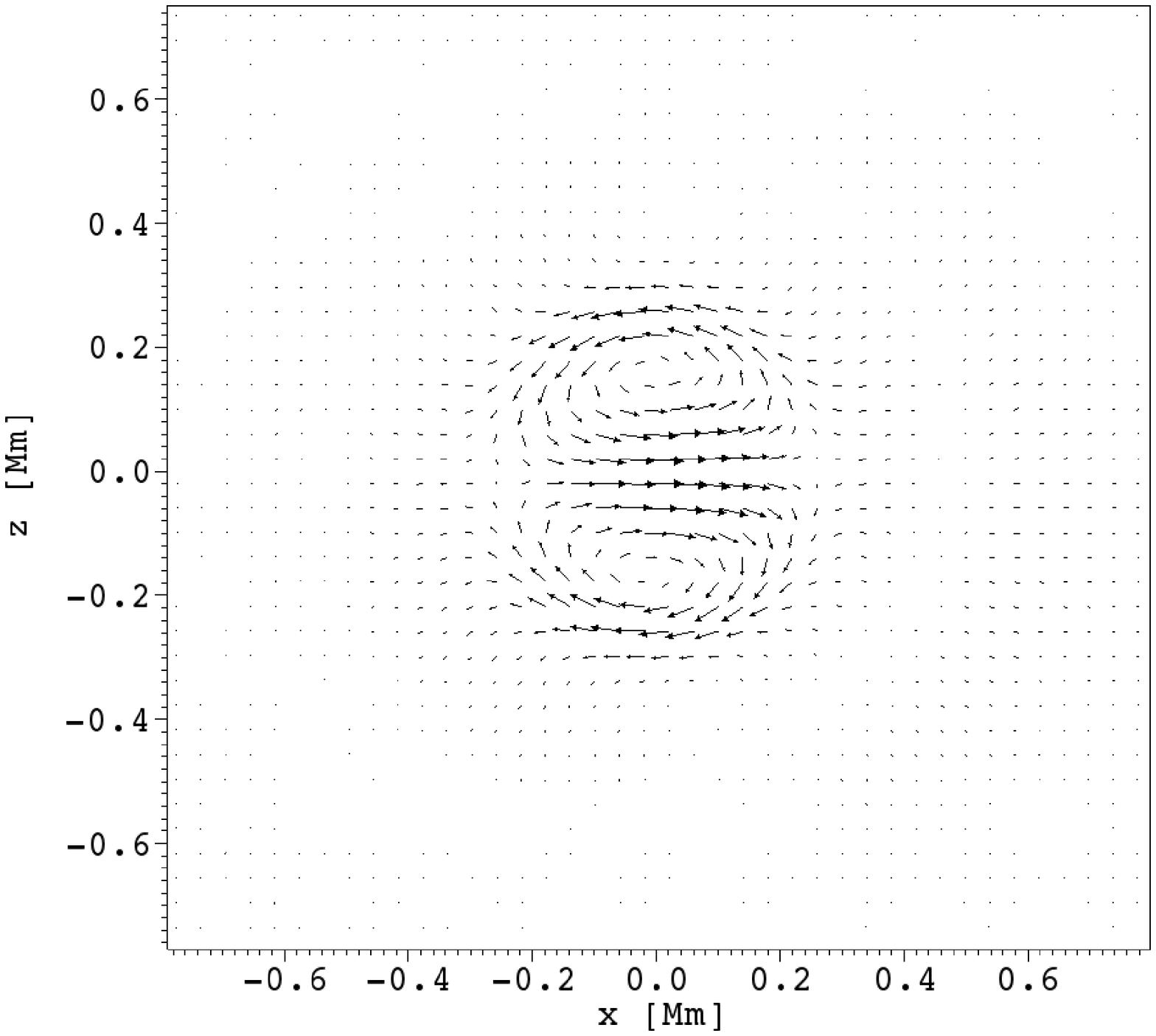}
\includegraphics[width=8.5cm]{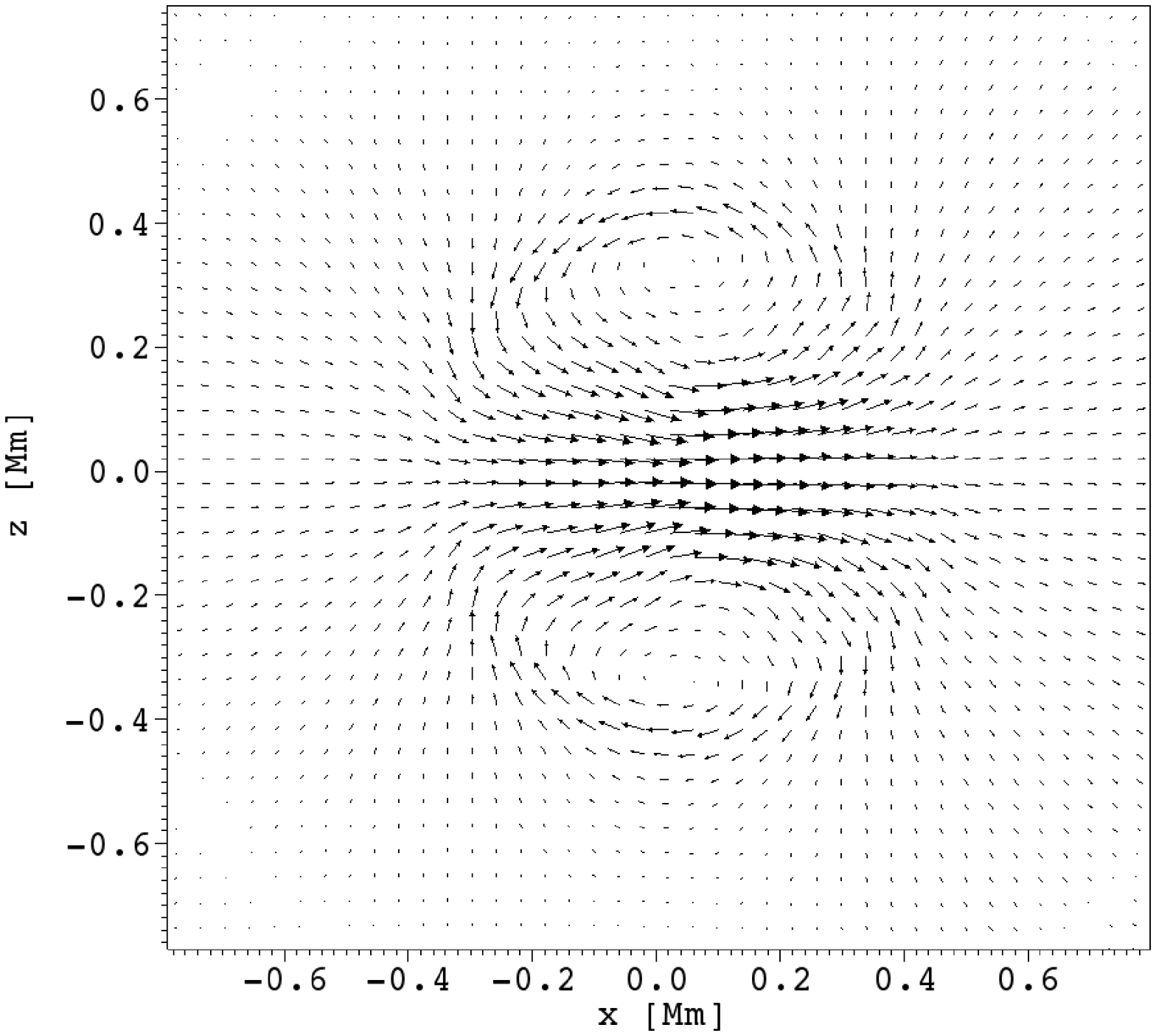}
\includegraphics[width=8.5cm]{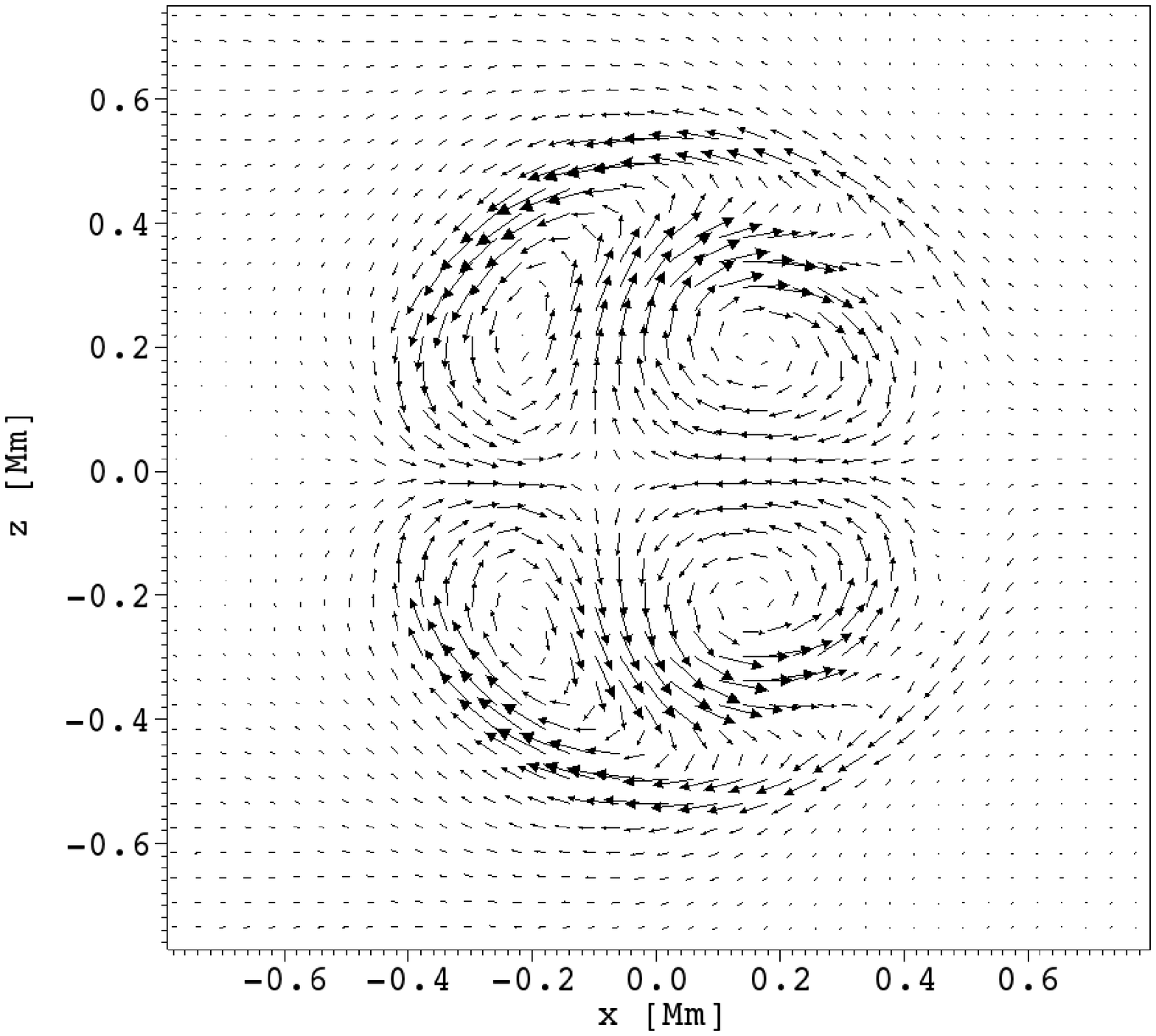}
\includegraphics[width=8.5cm]{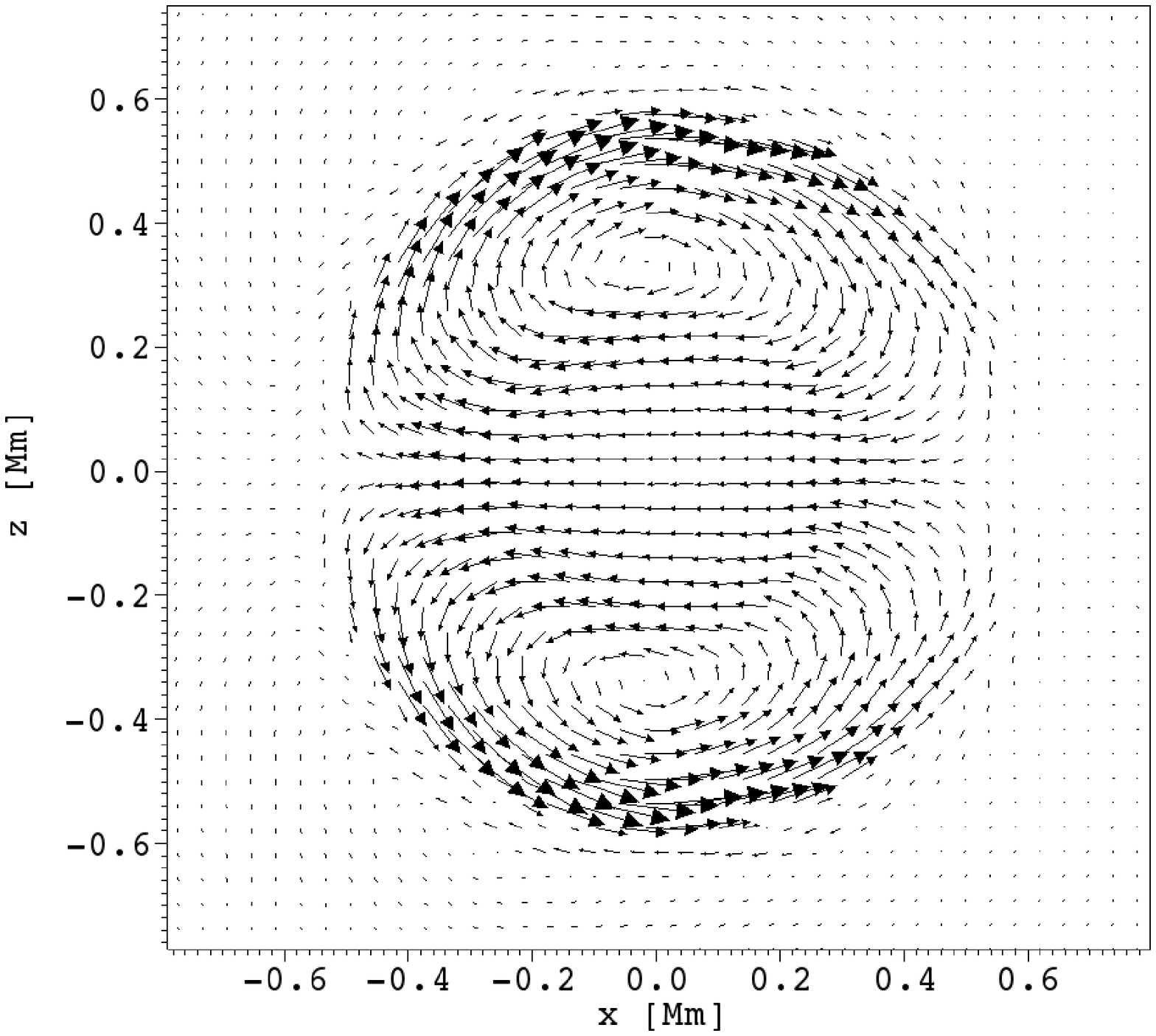}
\caption{Snapshots of velocity vectors (in arbitrary units) shown at $y=1.5$~Mm at times $t=125$~s (top left), $200$~s (top right), $300$~s (bottom left), and $550$~s (bottom right) for the case of the centrally-launched initial pulse.}
\label{fig:vert-V}
\end{figure*}
Figure~\ref{fig:vert-V} shows snapshots of the velocity vectors at several times after the initial pulse.
They are taken at the chromospheric altitude i.e. $y=1.5$~Mm.
The chromosphere is an ideal location to investigate the swirl structures in the solar atmosphere \citep[e.g.][]{Wed2012, Wed2013}.
Two confined vortices (i.e. swirl-like structures) are generated by the initial pulse.
The top swirl rotates anti-clockwise while the bottom swirl rotates clockwise, and the plasma flows reside within the area of the flux tube.

At $t=125$~s (top left panel of Fig.~\ref{fig:vert-V}) the swirls are mainly located at the center of the tube.
The dipolar flow pattern corresponds to a transverse motion of the central region as produced by a $m=1$ kink mode \citep[e.g. Fig. 3 of][]{Pascoe2010}.
At $t=300$~s (bottom left panel), four swirls (quadrupole type) are seen with the bottom left one rotating clockwise and the bottom right one rotating counter-clockwise, suggesting that the $m=2$ mode appears. 
The right pair of swirls result from the plasma reflection from the right side of the flux tube after 
the horizontally moving signal hits from the left side of the flux tube, and these reflected swirls have the opposite rotation to those which hit the right boundary.
At the same time, uprising swirling plasma motions are present at the left side.
Finally at $t=550$~s (bottom right panel), two well developed swirls with opposite rotating motions (top-swirl rotating clockwise and bottom-swirl 
rotating anti-clockwise) are seen.
It should be noted that these two swirls are settled on the outer edge of the flux tube, 
and the $m=1$ mode reappears. So, the number of modes varies in time from $m=1$ (Fig.~\ref{fig:vert-V}, top panels) 
to $m=2$ (bottom-left) and back to $m=1$ (bottom-right). 
%
\begin{figure*}
\centering
\includegraphics[width=8.5cm]{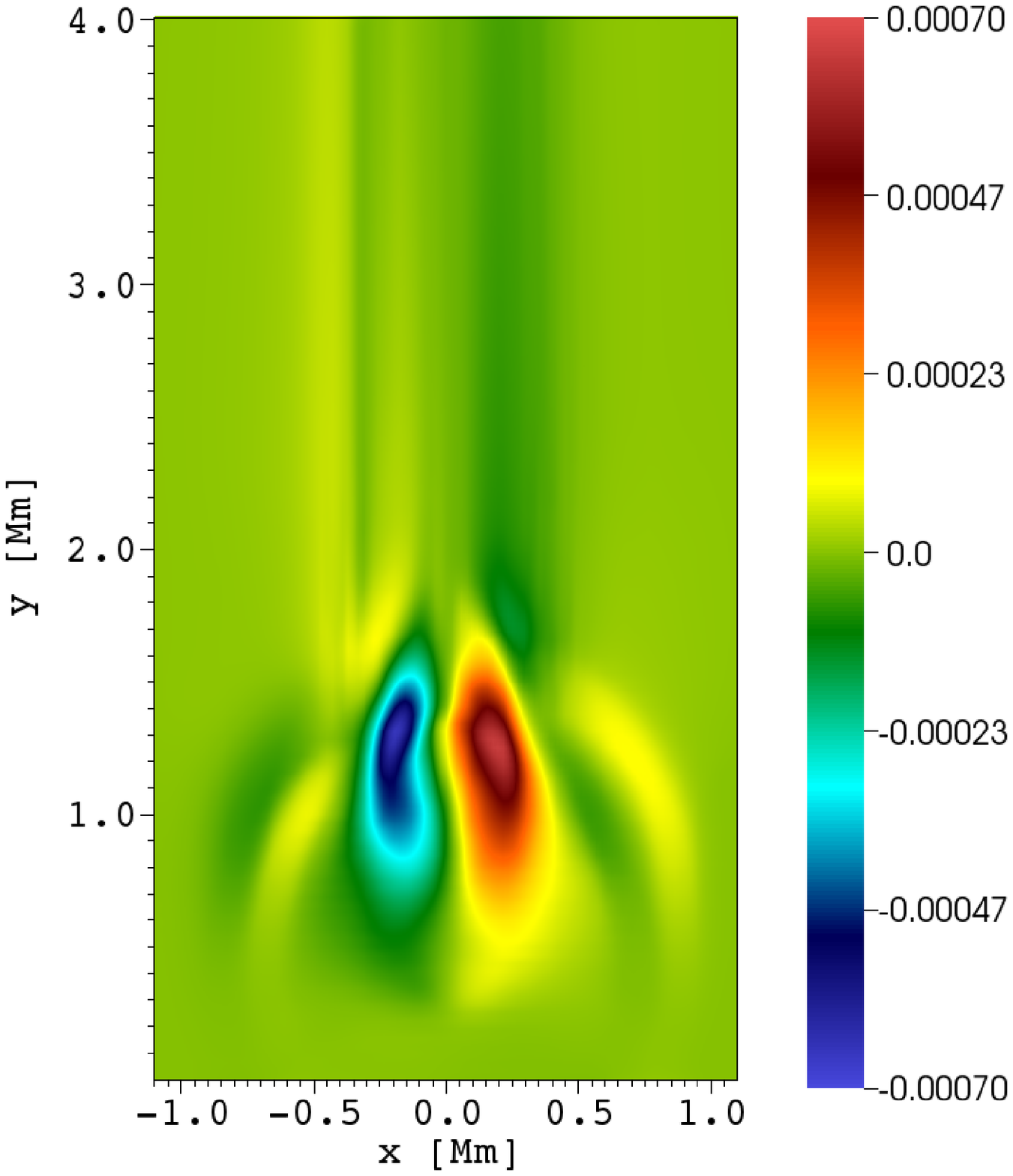}\hspace{-1.8cm}
\includegraphics[width=8.5cm]{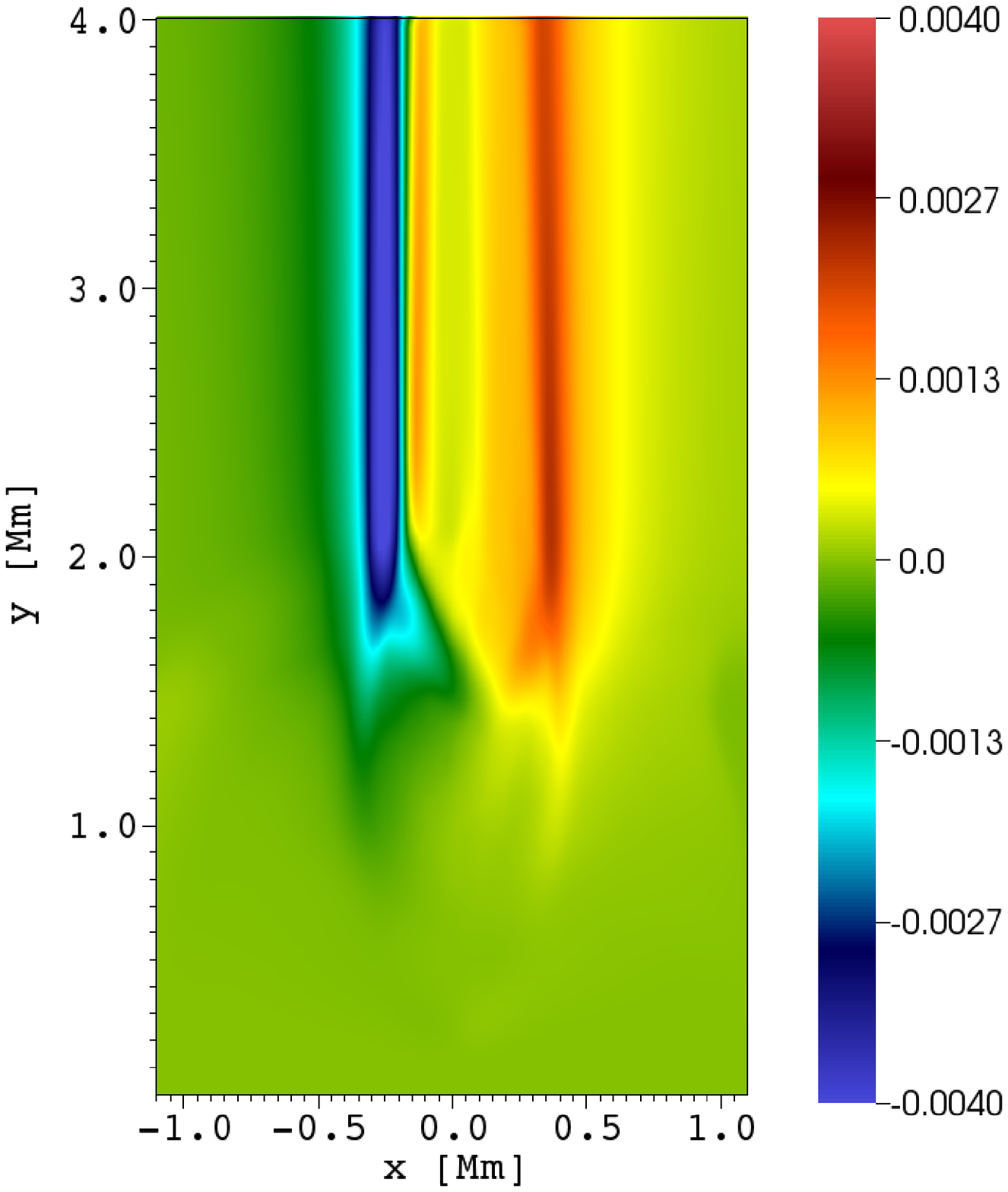}
\includegraphics[width=8.5cm]{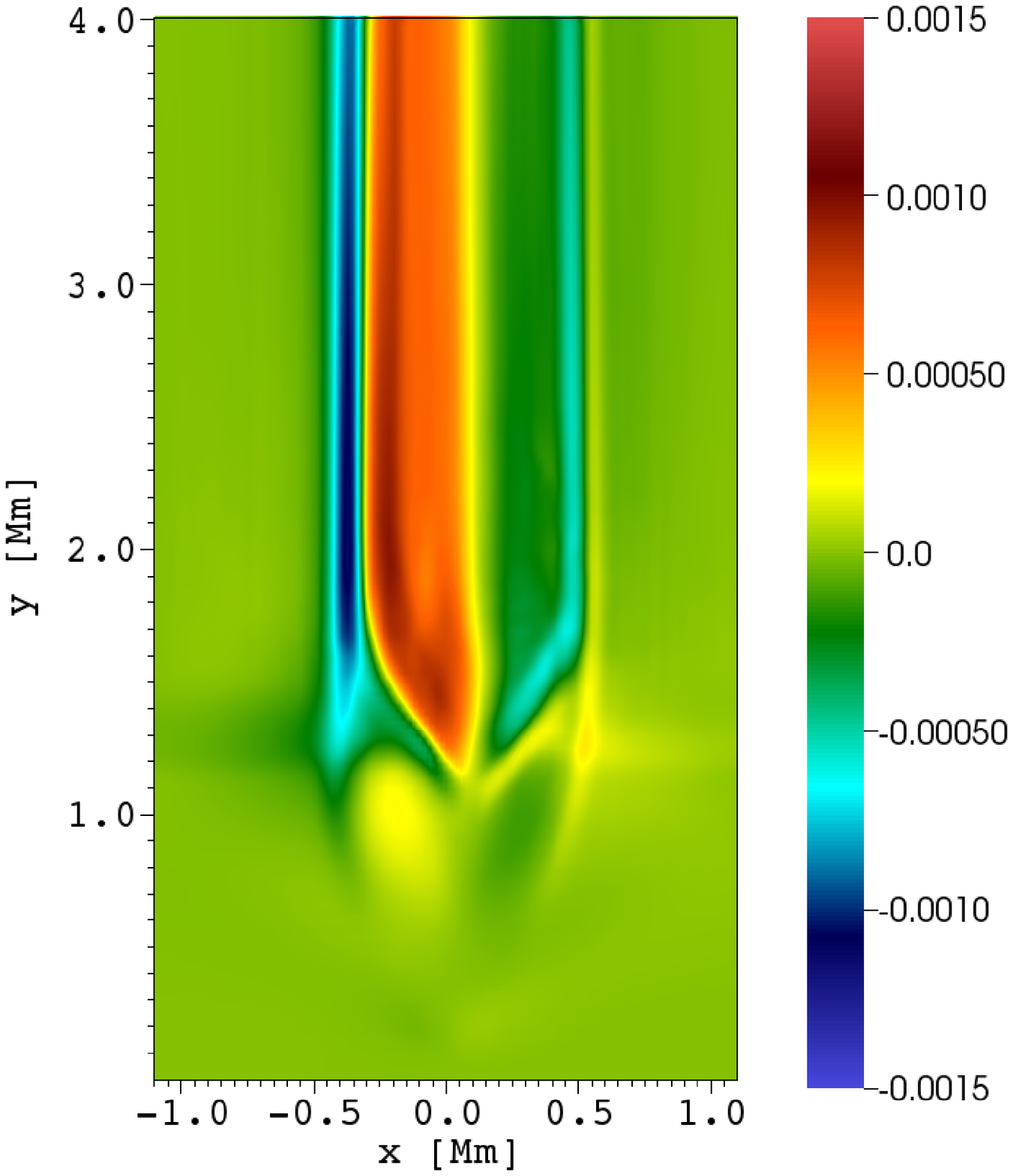}\hspace{-1.8cm}
\includegraphics[width=8.5cm]{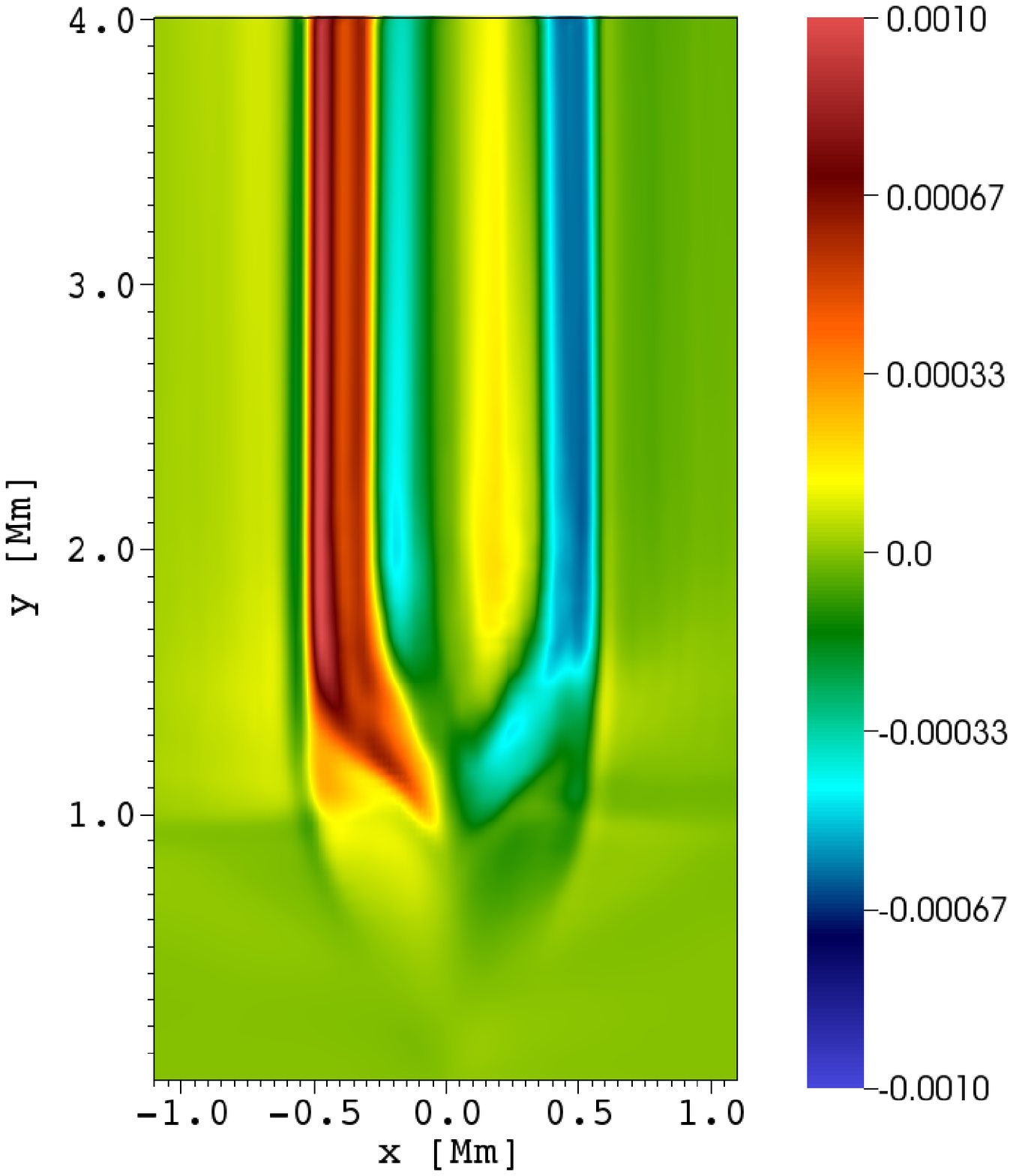}
\vspace{1.5cm}
\caption{Vertical profiles of $V_{\rm z}$ drawn at $z=0.2$~Mm at times $t=125$~s (top left), $200$~s (top right), $300$~s (bottom left), and $550$~s (bottom right) for the case of a centrally-launched initial pulse.
The color map corresponds to the magnitude and direction of velocity, expressed in units of $1$~Mm~s$^{-1}$.}
\label{fig:Vz_z=0.2}
\end{figure*}

Figure~\ref{fig:Vz_z=0.2} shows the spatial profiles of the transversal component of velocity, $V_{\rm z}$, which are displayed in the vertical plane, at $z=0.2$ Mm. 
The center of the flux tube (i.e. $z=0.0$~Mm) has not been chosen due to the presence of weak signal there due to the 
$m=1$ and $m=2$ symmetries of the oscillations.
The color map corresponds to the magnitude and direction of $V_{\rm z}$, with red and violet representing plasma moving towards and away from the observer, respectively.
Two regions of oppositely directed flow are visible at $t=125$~s (top left panel), moving away from the observer for $x < 0$, and towards the observer for $x > 0$.
This is a typical signature associated with rotating plasma as seen in terms of the transverse velocity and also a characteristic feature of the torsional Alfv{\'e}n waves as reported by \cite{Murawski2015b}.
At $t=200$~s, the perturbation has reached the solar corona and experienced horizontal expansion (top right panel).
The rotational flow grows in magnitude with time. At later times, plasma flows continue spreading in the horizontal direction with the presence of alternating segments of oppositely directed flow (bottom panels).
These alternating segments correspond to different magnetic interfaces (i.e. multi-shells) associated with Alfv{\'e}n waves, which are also reported by \cite{Murawski2015b}.
It should be noted that these velocity perturbations penetrate the solar corona and can transfer their energy into the solar corona, potentially fulfilling its energy losses.
%
\begin{figure*}
\centering
\mbox{
\hspace{-0.7cm}
\includegraphics[height=4.3cm]{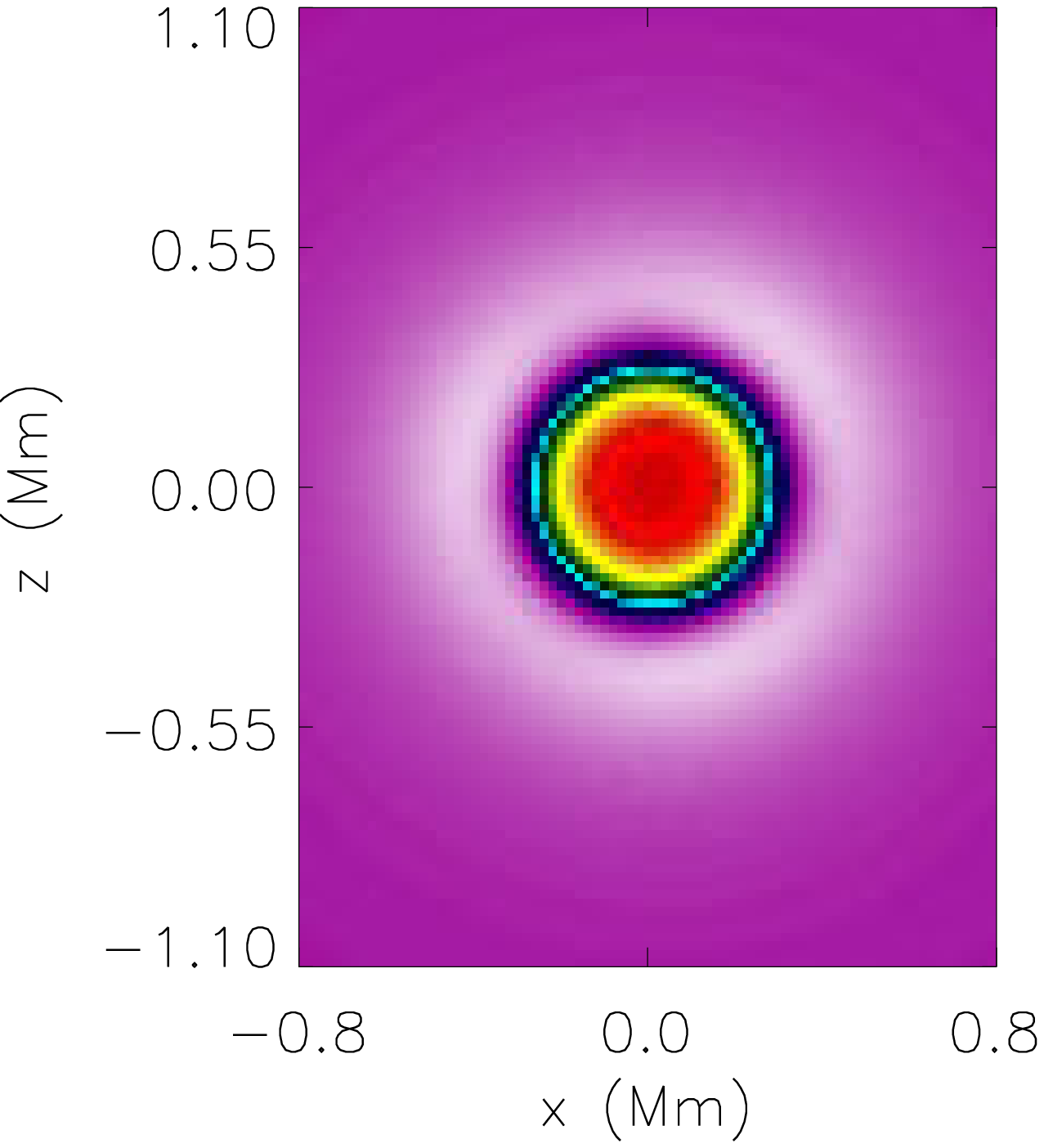}\hspace{-0.11cm}
\includegraphics[height=4.3cm]{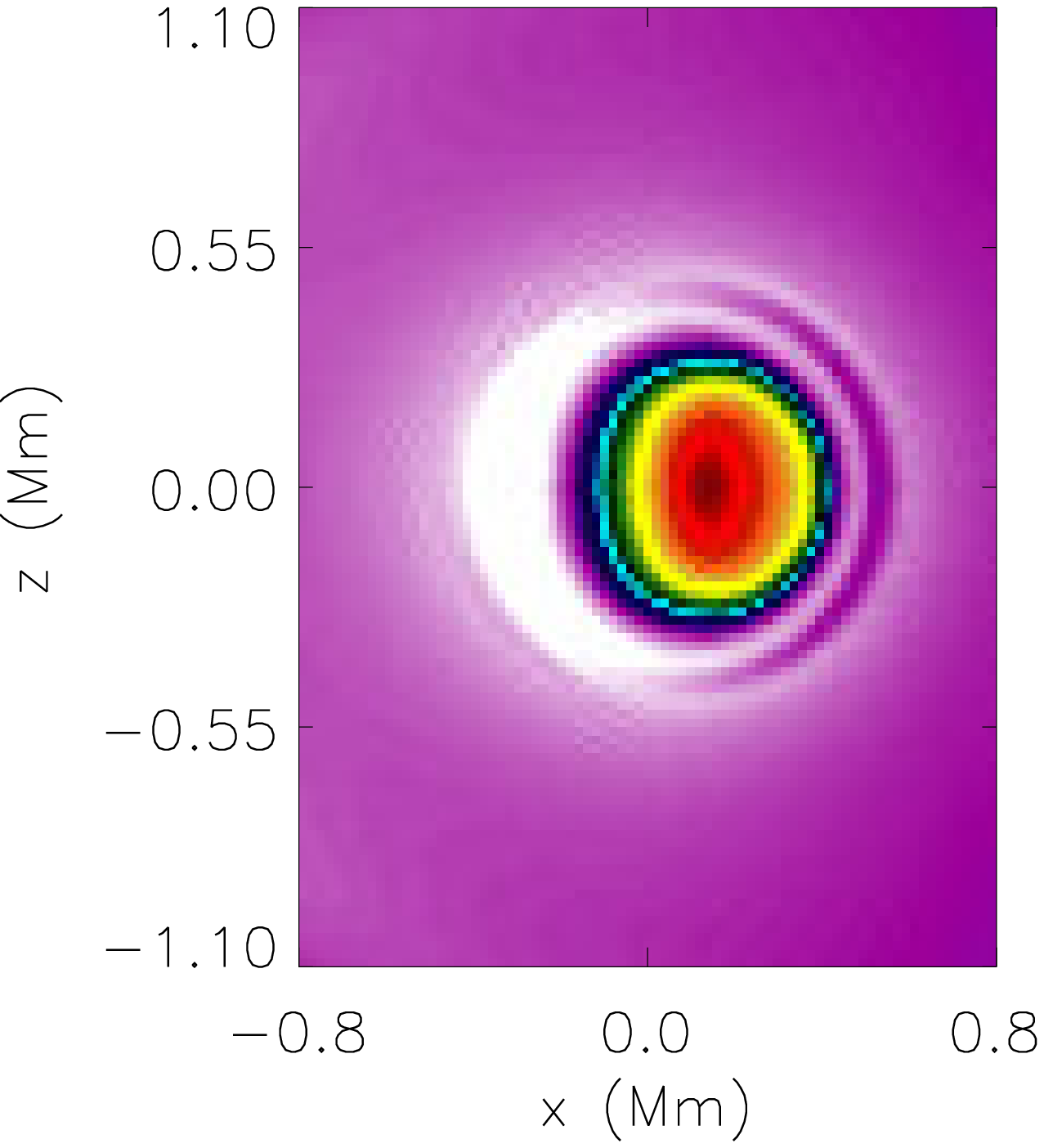}\hspace{-0.11cm}
\includegraphics[height=4.3cm]{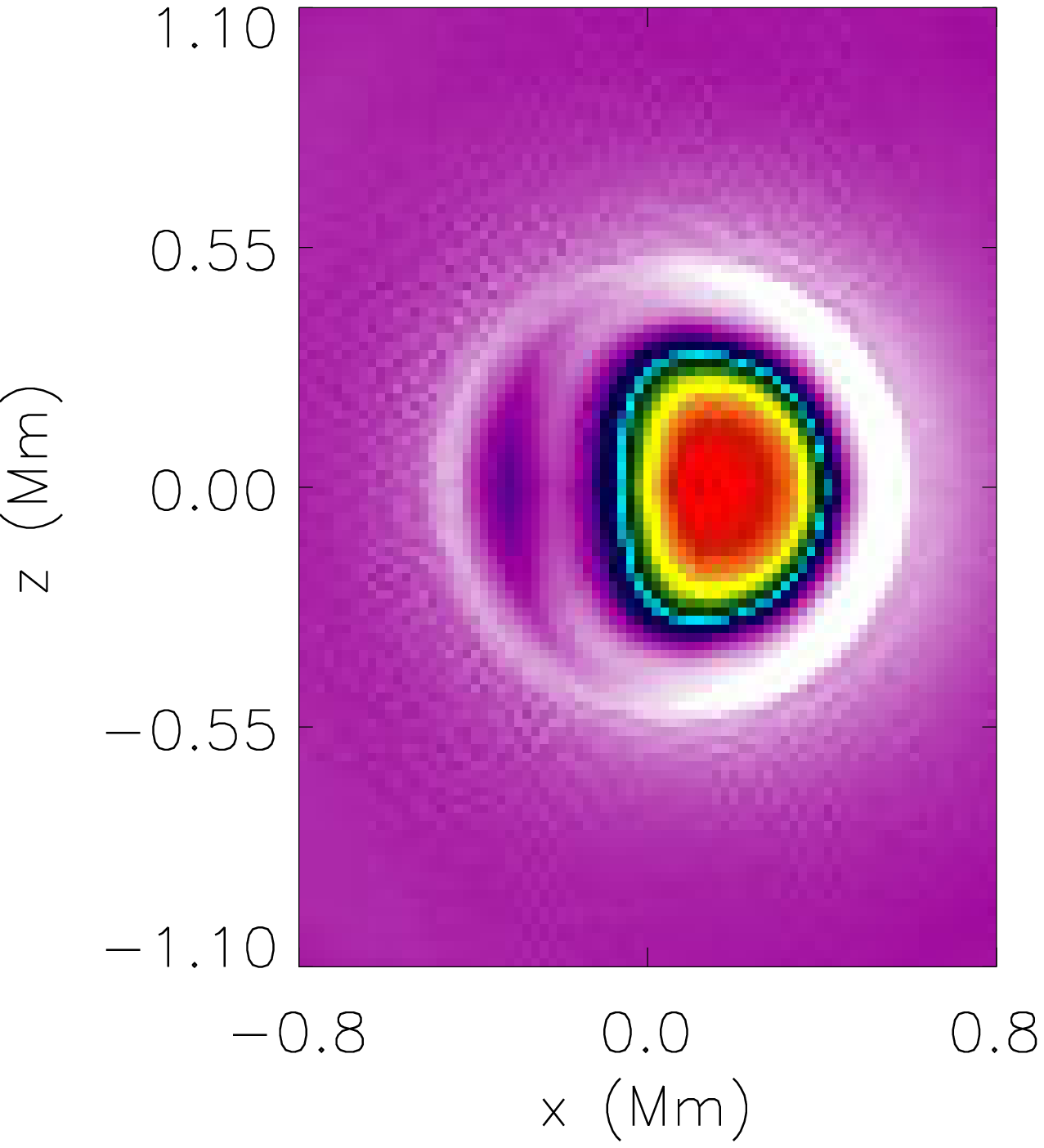}\hspace{-0.11cm}
\includegraphics[height=4.3cm]{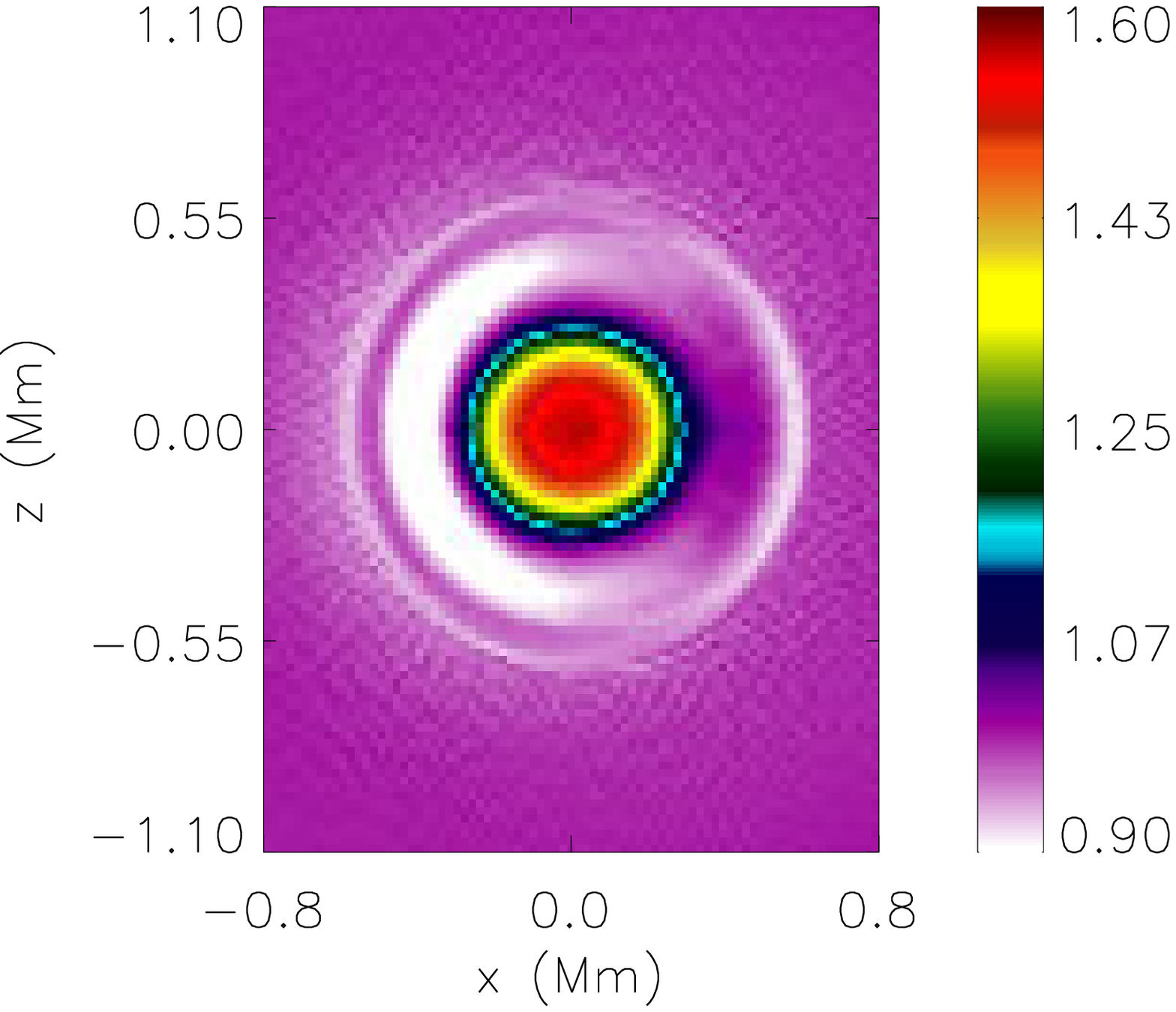}
}
\mbox{
\includegraphics[height=4.6cm]{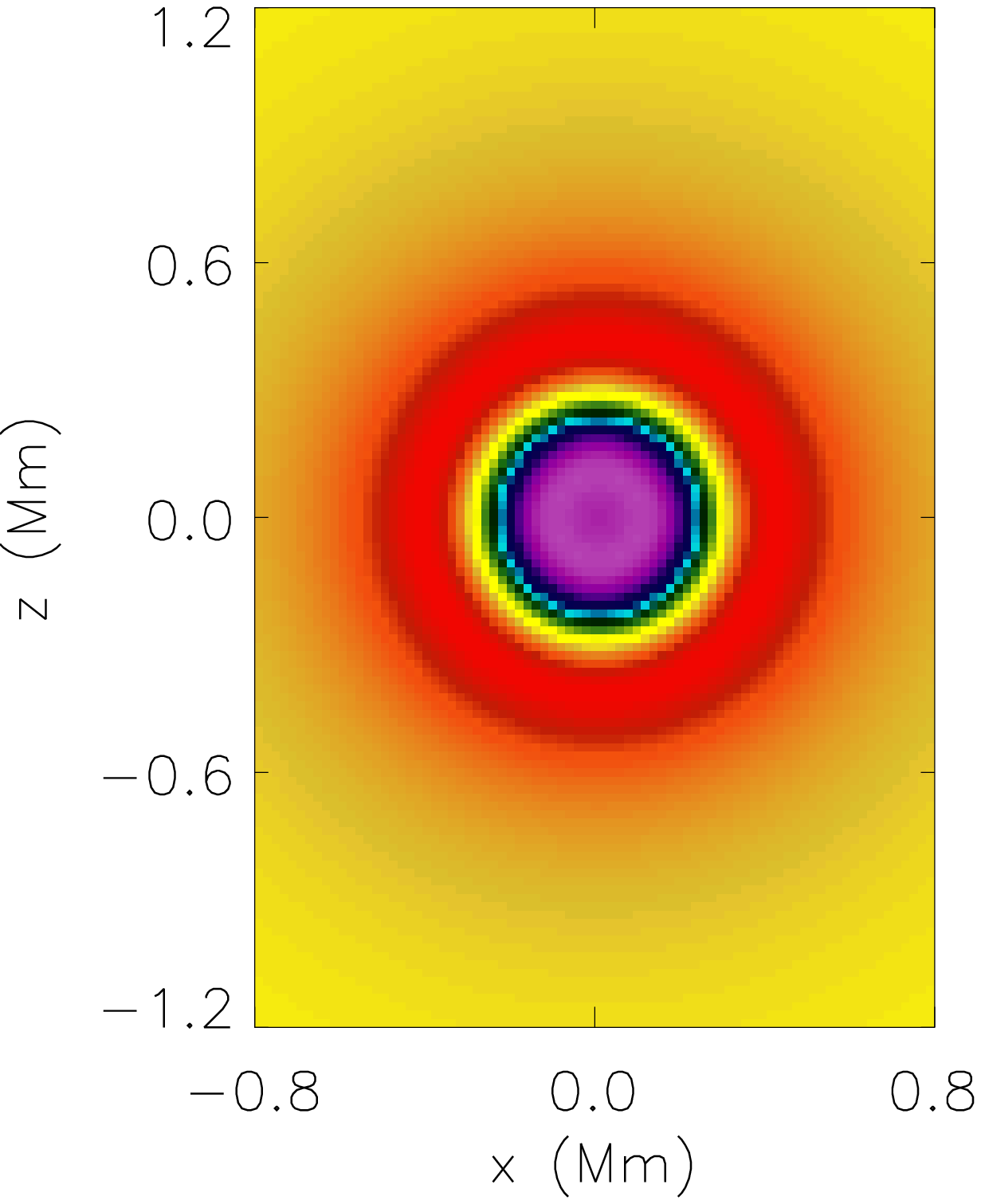}
\includegraphics[height=4.6cm]{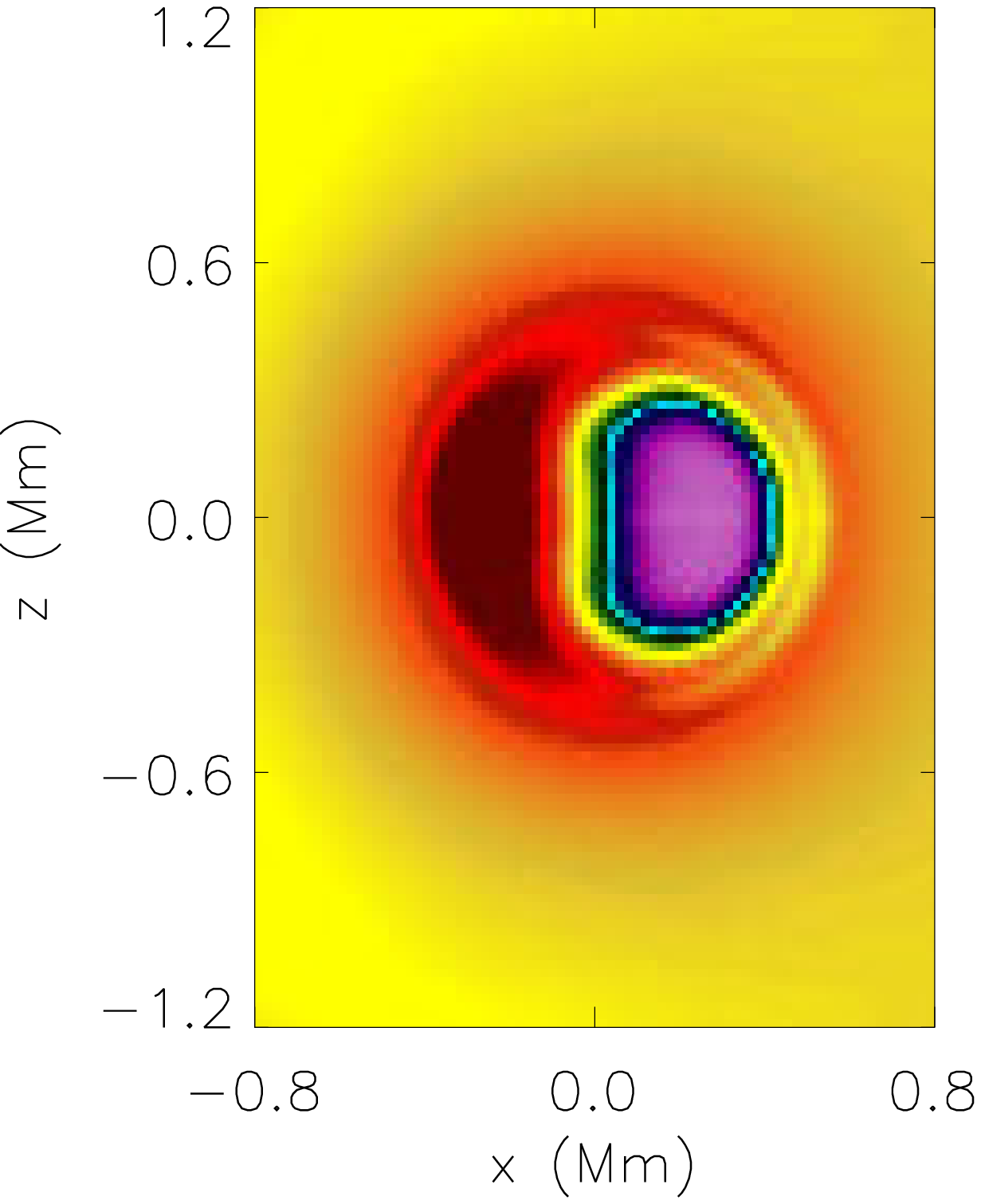}
\includegraphics[height=4.6cm]{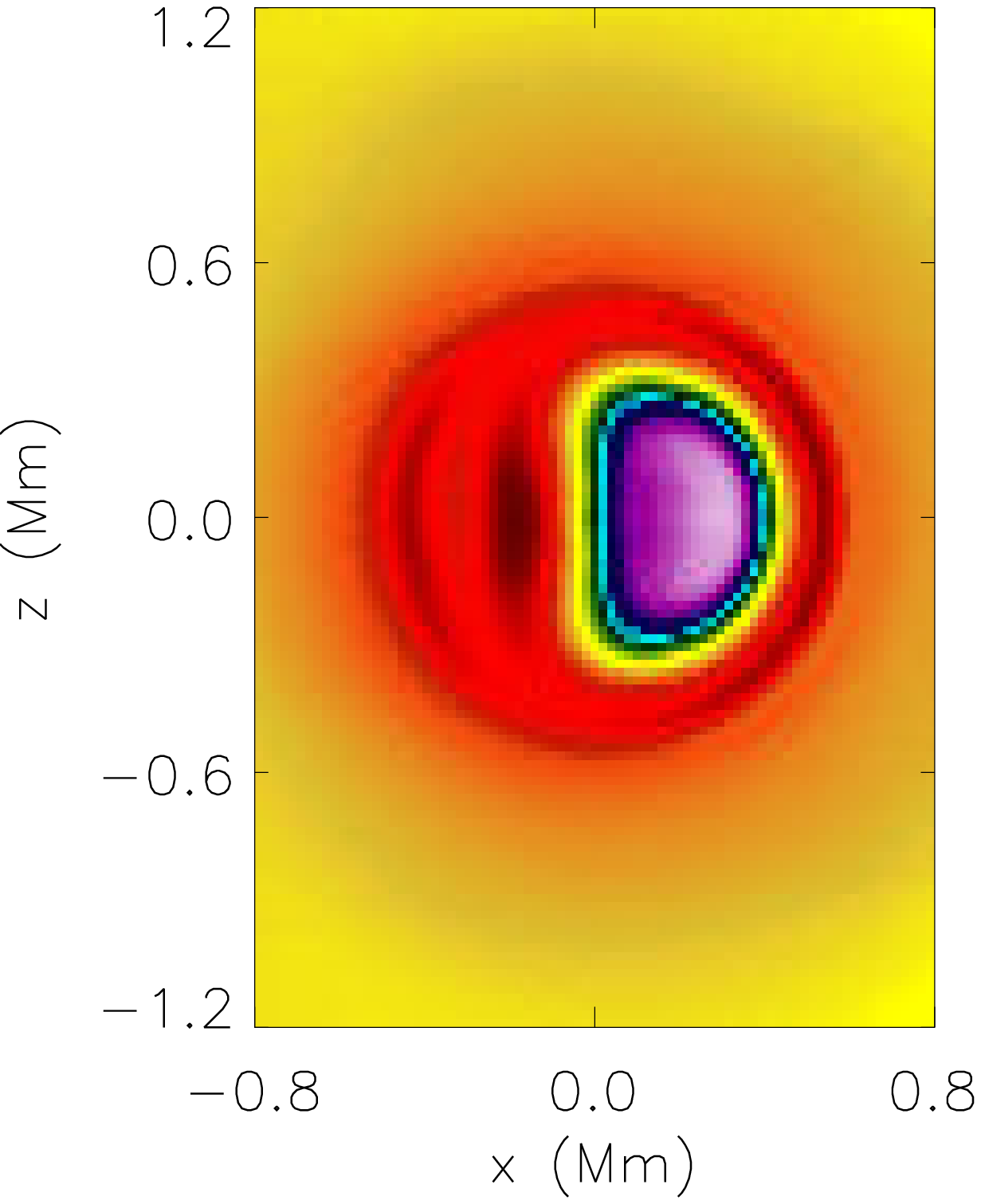}
\includegraphics[height=4.6cm]{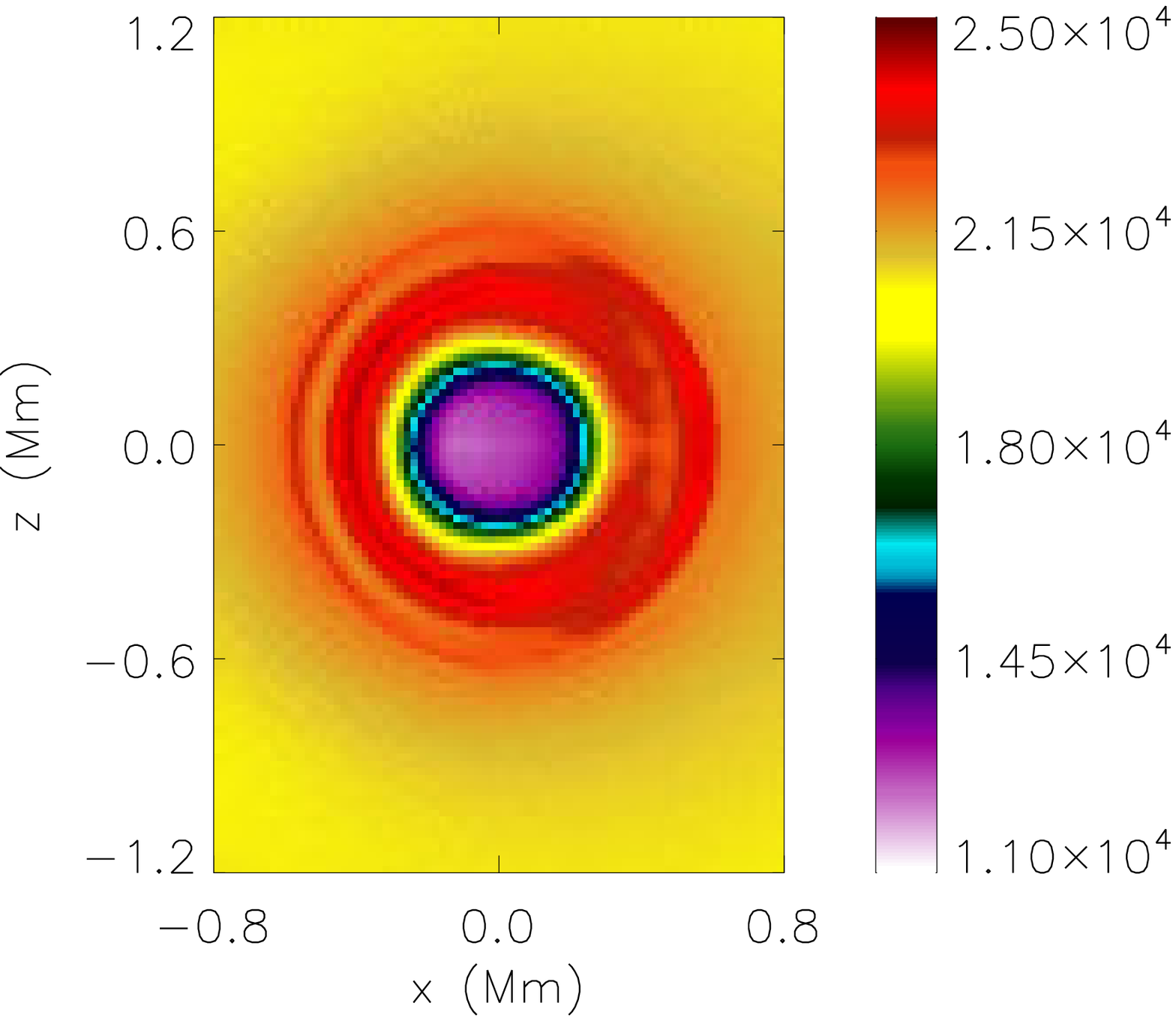}
}
\caption{Horizontal profiles of $B_{\rm y}$ (top row) and mass density $\varrho$ (bottom row) at $y=1.5$~Mm at times (left to right) $t=0$, $250$, $350$, and $600$~s.
The color maps corresponds to the magnitude of a magnetic field expressed in units of $\approx 11.4$~G (top row), and the magnitude of $\varrho$ expressed in units of $10^{-12}$~kg~m$^{-3}$ (bottom row).}
\label{fig:vert-B}
\end{figure*}

It is noteworthy that centrally-launched initial pulse alters the magnetic field lines of the flux tube. 
We show the vertical magnetic field component at a height of $y=1.5$ Mm at four different moments of time (top panels of Fig.~\ref{fig:vert-B}).
The color map represents the strength of the magnetic field, using a logarithmic scale. 
The initial state is displayed at $t=0$~s (top left panel), which shows a high concentration of magnetic field at the center of the flux tube and decreasing towards the outer edges of the flux tube.
The profile at $t=0$~s shows that the diameter of the flux tube is approximately $0.6$~Mm. 
At $t=250$~s, as a result of the initial pulse, the magnetic field lines are displaced towards the right side from the center of the flux tube.
At $t=350$~s, the magnetic field begins to return towards the left side which is evidenced by the white patch of a weaker $B_{y}$ seen at the right side of the flux tube.
At $t=600$~s, the profile of $B_{y} (x,y=1.5,z)$ is similar to that at $t=0$~s, indicating that the flux tube moves back through its equilibrium state.
This temporal evolution of the horizontal profiles of the vertical component of magnetic field shows that the magnetic field lines are initially displaced from the central axis $x=0$~Mm, $z=0$~Mm 
and then oscillate around their equilibrium.
In addition to the temporal evolution of horizontal magnetic field, we also investigate the horizontal profiles of mass density at different times (bottom panels of Fig.~\ref{fig:vert-B}).
The temporal evolution of $\varrho (x,y=1.5,z)$ closely follows the evolution of vertical component of magnetic field shown in the top panels.
The mass density profiles are altered at later moments of time in the form of plasma compressions and rarefactions which occur alternatively at the left and right sides of the flux tube.
Such compressions and rarefactions in the flux tube correspond to fast magnetoacoustic kink waves (kink, henceforth).
Therefore, we can say that these collective signatures (i.e. oscillations of horizontal magnetic field and mass density) indicate the presence of kink waves.
We note developed dense concentric shells which are discernible for $t\ge 350$~s.
These magnetic shells become most pronounced at $t=600$~s (bottom right panel of Fig.~\ref{fig:vert-B}), and result from up-flowing and down-flowing plasma; the upwards propagating plasma brings dense gas from lower atmospheric regions, while downflows leads to plasma evacuation which are well seen in the form of bows.
Hence we obtain complex dynamics of the flux tube in response to the centrally-launched pulse.
Double magnetic swirls (i.e. eddies), Alfv{\'e}n perturbations at various magnetic shells and related transverse motions (Fig.~\ref{fig:vert-B}), 
and rotational flows in the flux tube (Fig.~\ref{fig:vert-V}) are clearly evident.
It should be noted that similar vortices were also reported by \cite{Murawski2015} in the case of horizontally homogeneous solar atmosphere, 
and vortices were found to accompany kink waves which couple to the $m=1$ Alfv\'en waves in flux tubes embedded in a gravity-free and cold plasma, arising due to the resonant absorption of kink modes \citep[compare Fig.~12 of][with the bottom right panel of Fig.~\ref{fig:vert-V}]{Goossens2014}.
\citet{Pascoe2011} demonstrated that resonant absorption efficiently couples kink and Alfv\'en modes in an arbitrary inhomogeneous medium and that decoupled Alfv\'en waves could also be established.
The direct and indirect excitation of Alfv\'en waves by footpoint motions was studied also by \cite{Goossens2001}.
\citet{2012ApJ...746...31D} discussed the ambiguity of Doppler observations for mode identification, particularly in the optically thin corona which allows line-of-sight integration of multiple structures.
Our equilibrium is also similar to the (2D) funnel considered by \citet{Pascoe2014}, 
for which the flux tube is an anti-waveguide for fast magnetohydrodynamic waves and so the kink waves will also propagate away from the central region. 
This effect will also contribute to the appearance of the concentric shells and the Alfv\'en waves becoming stronger at later times.
\begin{figure*}
\begin{center}
\includegraphics[width=5.8cm]{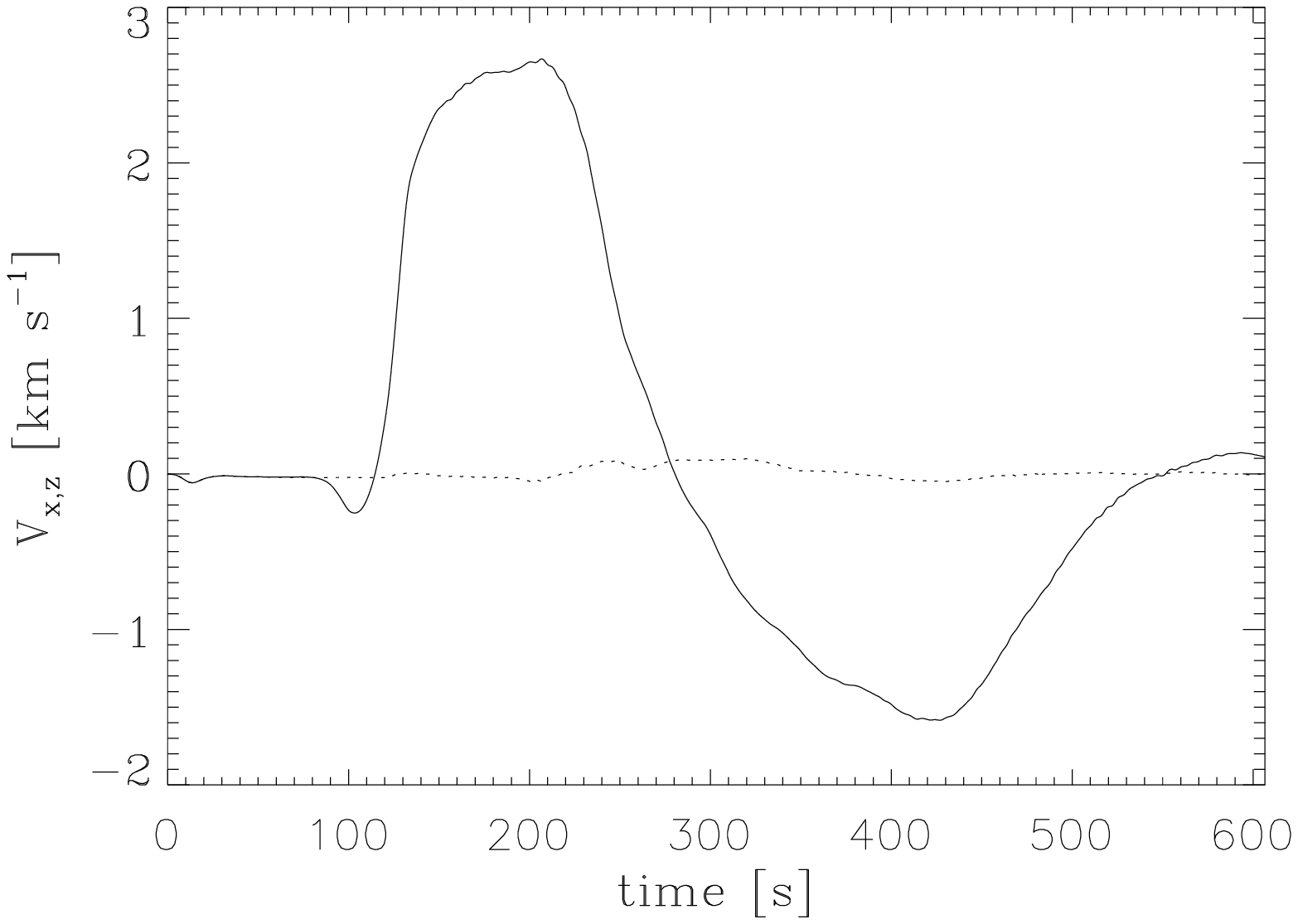}
\includegraphics[width=5.8cm]{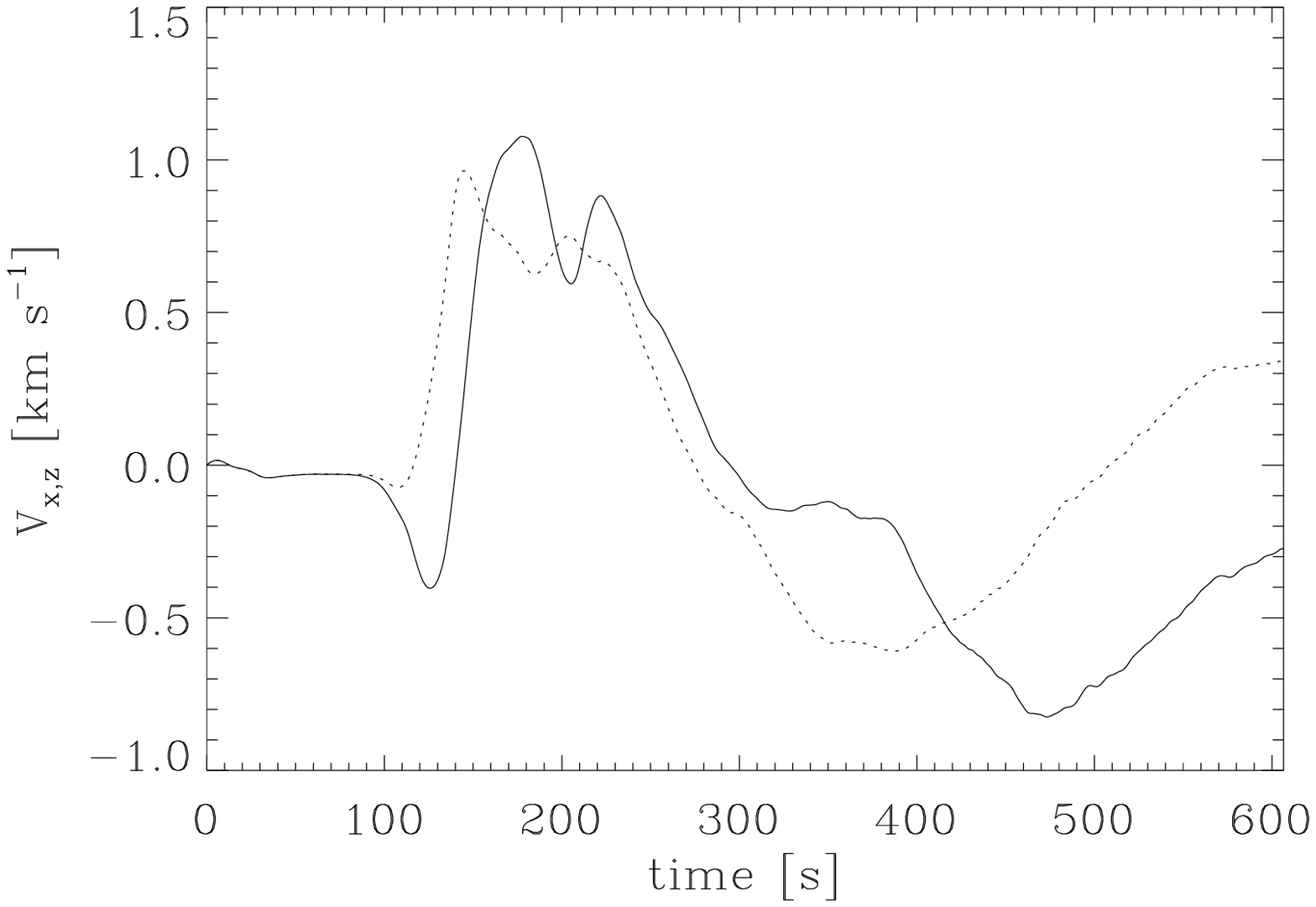}
\includegraphics[width=5.8cm]{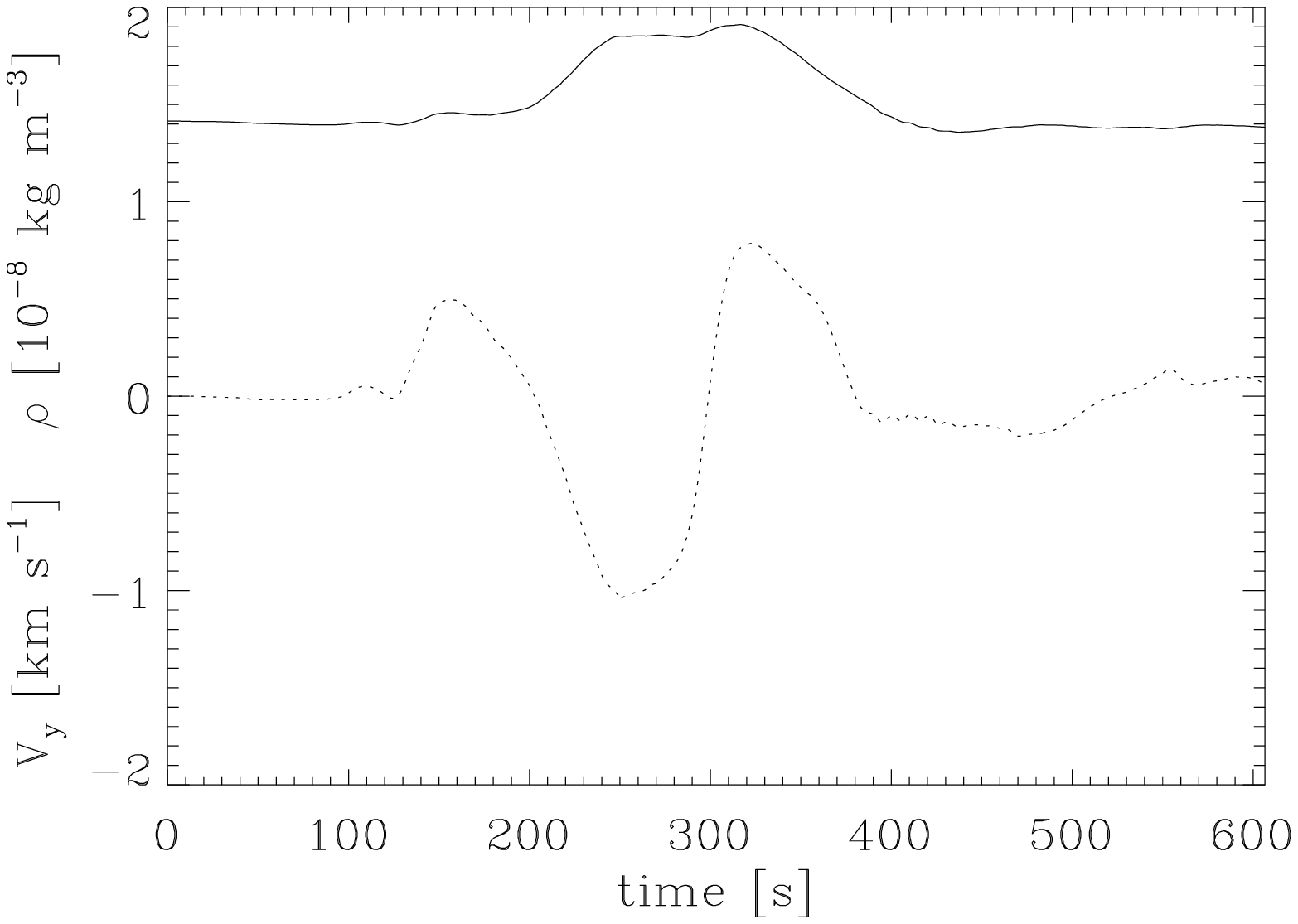}
\caption{Time signatures created by a centrally-launched pulse for: (a) left panel: $V_{\rm x}$ (solid line) and $V_{\rm z}$ (dotted line) measured at the detection point $(x=0, y=1.5,z=0)$~Mm;
(b) middle panel: $V_{\rm x}$ (solid line) and $V_{\rm z}$ (dotted line) measured at $(x=0.2, y=1.5,z=0.2)$~Mm (middle); 
(c) right panel: $\varrho$ (solid line) and $V_{\rm y}$ (dotted line) measured at $(x=0, y=1.5,z=0)$~Mm.}
\label{fig:ts}
\end{center}
\end{figure*}

Figure~\ref{fig:ts} shows time signatures of $V_{\rm x}$ and $V_{\rm z}$ measured at different detection points (left and middle panels).
These signatures reveal perturbations in which $V_{\rm x}$ dominates over $V_{\rm z}$ at the flux tube center (left panel), 
but on the side of the tube, $V_{\rm z}$ is of a comparable magnitude to $V_{\rm x}$ (middle panel).
The right panel illustrates $\varrho$ (solid line) and $V_{\rm y}$ (dotted line), measured at $(x=0,y=1.5,z=0)$~Mm.
We note that $\varrho$ and $V_{\rm y}$ are 
essentially correlated, being a sign of slow magnetoacoustic waves and showing that upflow brings dense plasma from lower layers, while downwardly propagating plasma reduces the mass density.

Hence we have shown the presence of Alfv{\'e}n and kink waves as well as slow magnetoacoustic waves in the flux tube.
Alfv{\'e}n waves propagate into the solar corona (see Fig.~\ref{fig:Vz_z=0.2}).
The flux tube used in our simulations consists a region of enhanced Alfv{\'e}n speed and therefore it is an anti-waveguide for kink waves, which are attenuated by leaking their energy to the ambient medium.
However, the kink waves still play a significant role in distribution of energy into the upper atmosphere.
\begin{figure*}
\centering
\includegraphics[height=7.1cm]{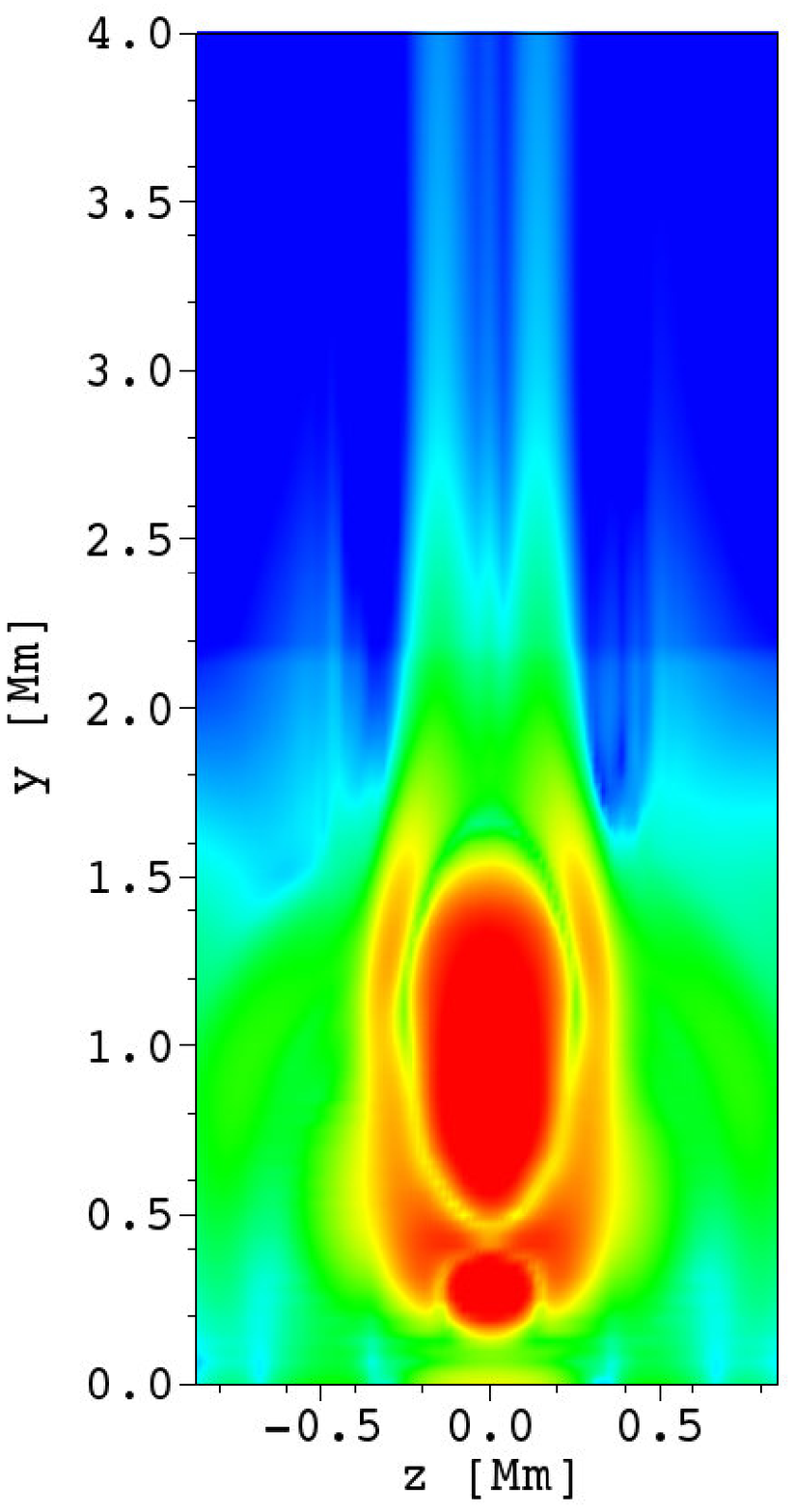}
\includegraphics[height=7.1cm]{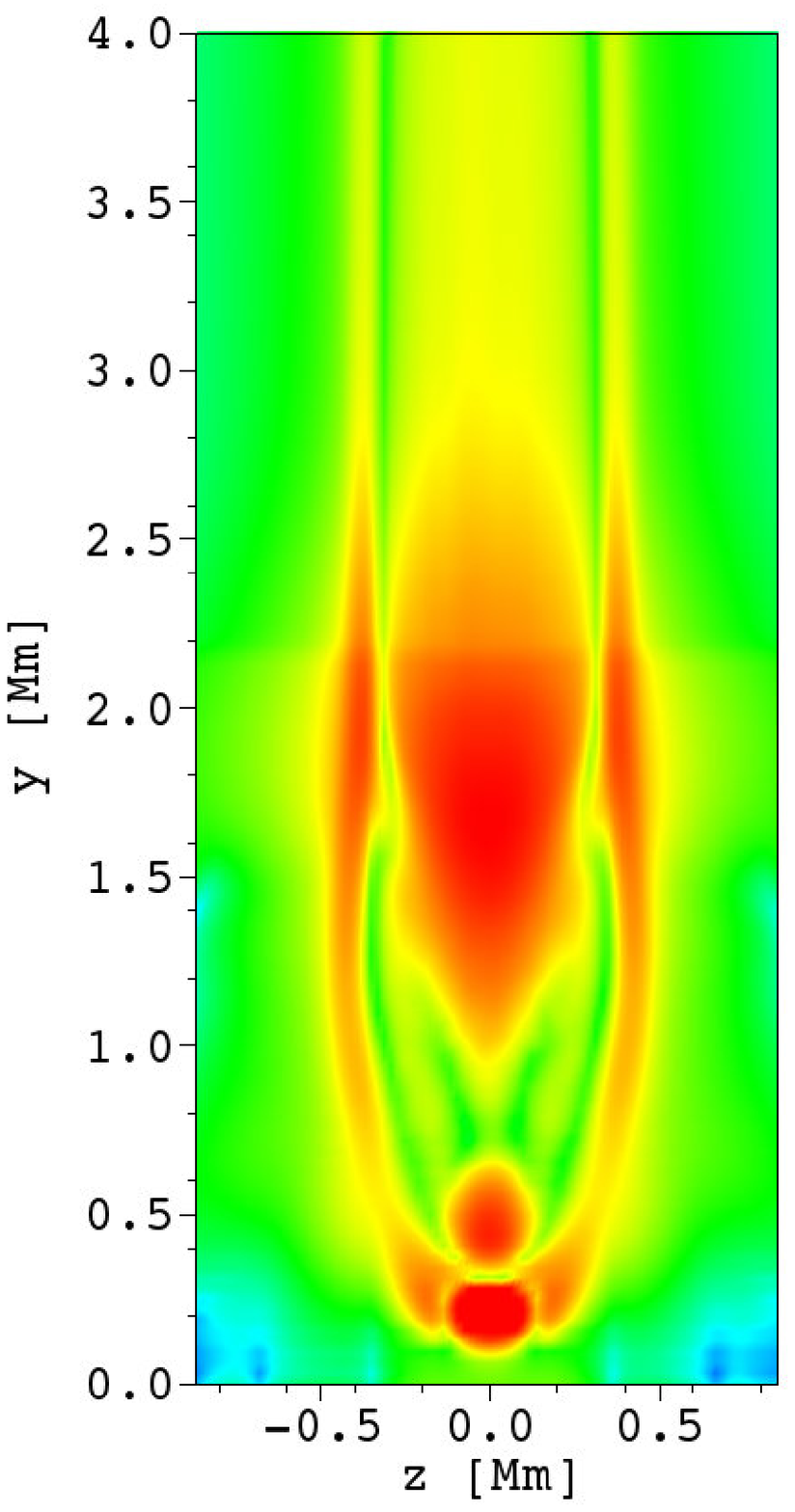}
\includegraphics[height=7.1cm]{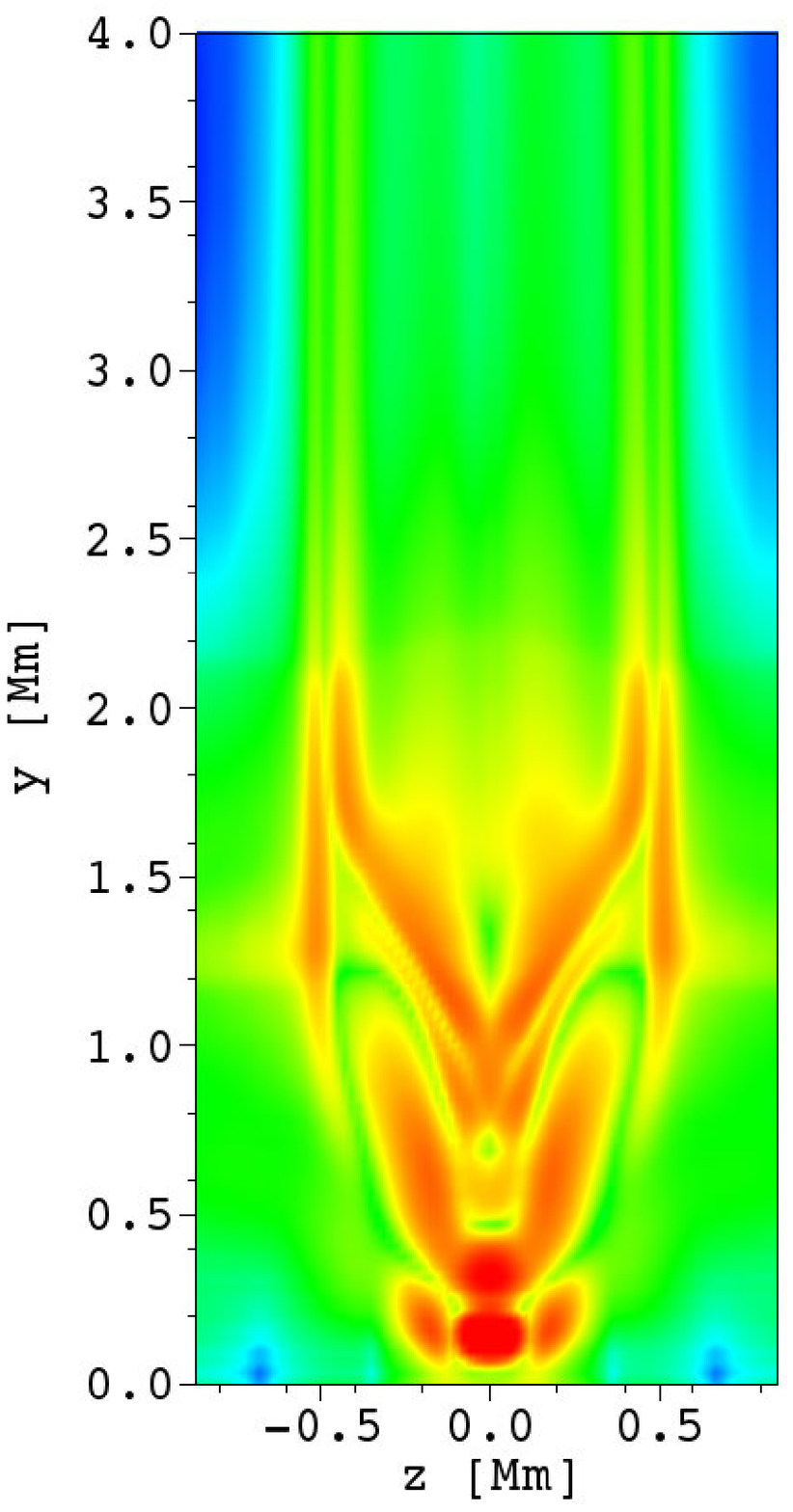}
\includegraphics[height=7.1cm]{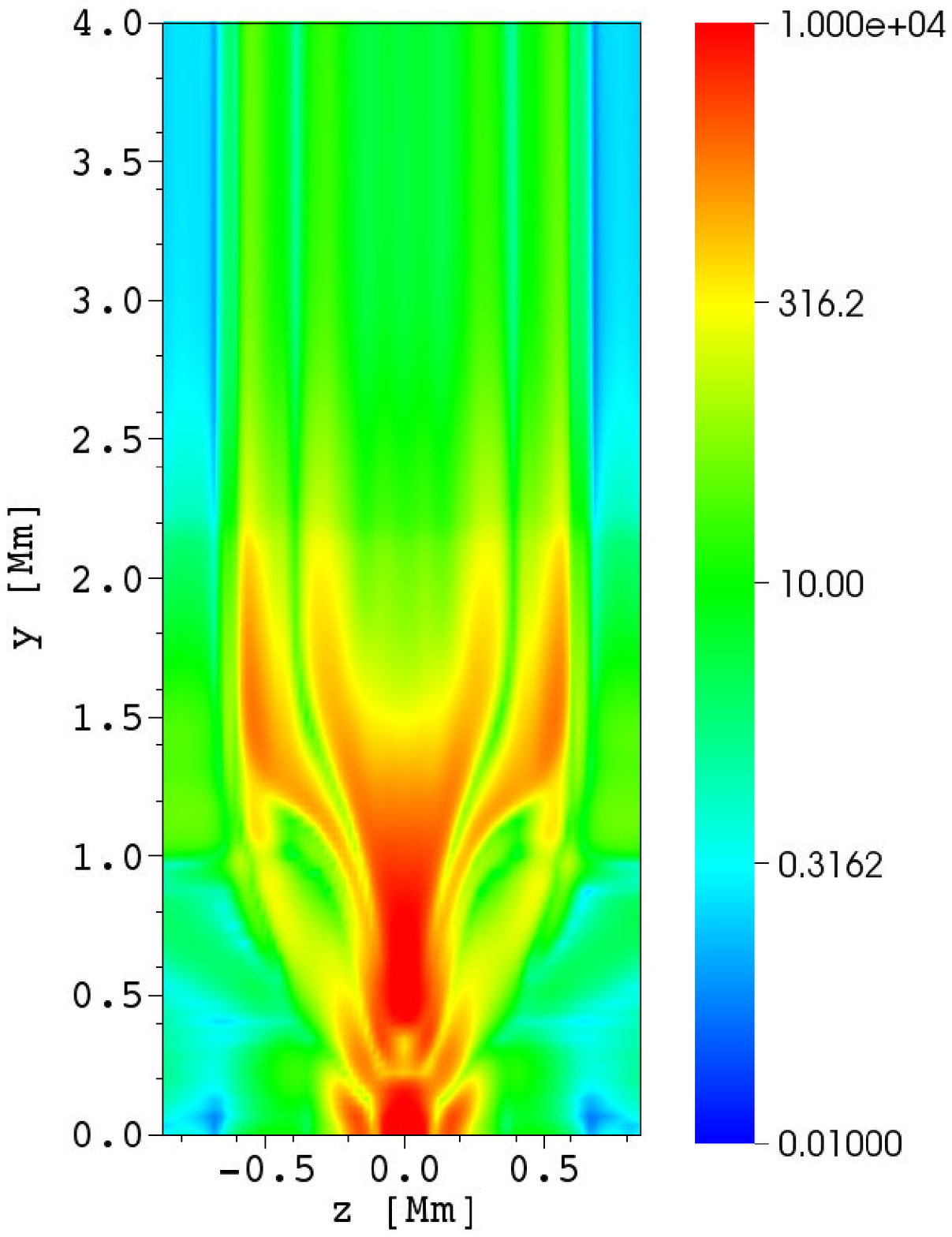}
\caption{Temporal evolution of the energy flux,
$F \approx 0.5 \varrho_{0} c_{\rm A} V^{2}$,
expressed in units of W~m$^{-2}$, drawn at $x=0$~Mm at times (left to right) of $t=125$, $200$, $300$, and $550$~s for the case of a centrally launched initial pulse.}
\label{fig:fluxcentral}
\end{figure*}

Figure~\ref{fig:fluxcentral} shows the estimated energy flux carried into the upper atmosphere.
During the evolution of the double vortex associated with $m=1$ kink waves up to $t=200$~s, 
the amount of energy filling the chromosphere $\sim 10^{4}$~W~m$^{-2}$ is sufficient to fulfill its large losses due to radiation and mass transport.
The initial wave transporting the energy to the inner corona ($y=2.5$~Mm) is also sufficient to fulfill the coronal energy losses ($\sim 10^{2}$~W~m$^{-2}$).
With the development of more complicated vortex patterns at later times, 
a sufficient amount of energy is still present at the chromospheric heights below the transition-region ($y=2.1$ Mm). 
However, the energy is spread more towards the boundary of the flux tube, and the amount of energy being channelled to the corona is reduced.
Therefore, the initial evolution of the vortex motion is an important phase in channelling the energy from the chromosphere to the corona. 
However, the energy flux at the bottom of the corona is $3 \times 10^{2}$~W~m$^{-2}$, which is still sufficient to fulfill its loses. 
Considering that there are no dissipation mechanisms in the model considered, the energy transported seems to be directly related 
to the amplitude of the pulse driver. 
However, the solar chromosphere is dominated by collisions and it is partially ionized. 
In this framework, the waves can be dissipated earlier in the chromosphere by such non-ideal effects, e.g., magnetic diffusivity, plasma viscosity, ion-neutral drag forces, and ambipolar diffusion. 
Significant parts of the waves are also efficiently reflected by the gradients in the sound and Alfv\'en speeds at the transition region. 
Thus, the energy flux deposited in the corona is likely an upper bound in the present model. 
\subsection{Perturbation by an off-center pulse}
Here we investigate the dynamics of the flux tube in response to an off-center pulse applied in the horizontal component of velocity $V_{x}$.
This pulse is described by Eq.~(\ref{eq:perturb}) with $x_{\rm 0}=0$~Mm and $z_{\rm 0}=0.3$~Mm.
\begin{figure*}
\centering
\includegraphics[width=8.5cm]{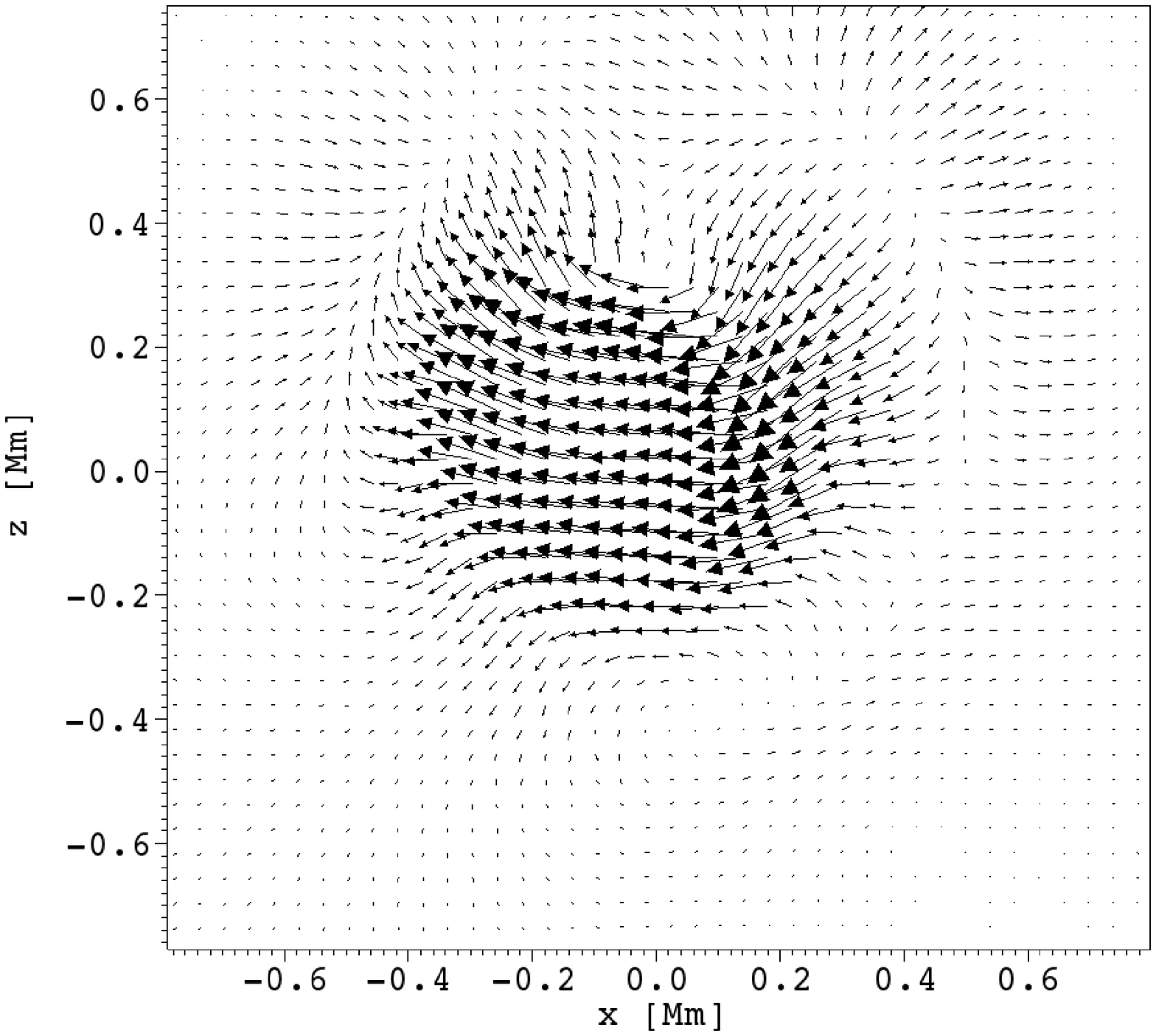}
\includegraphics[width=8.5cm]{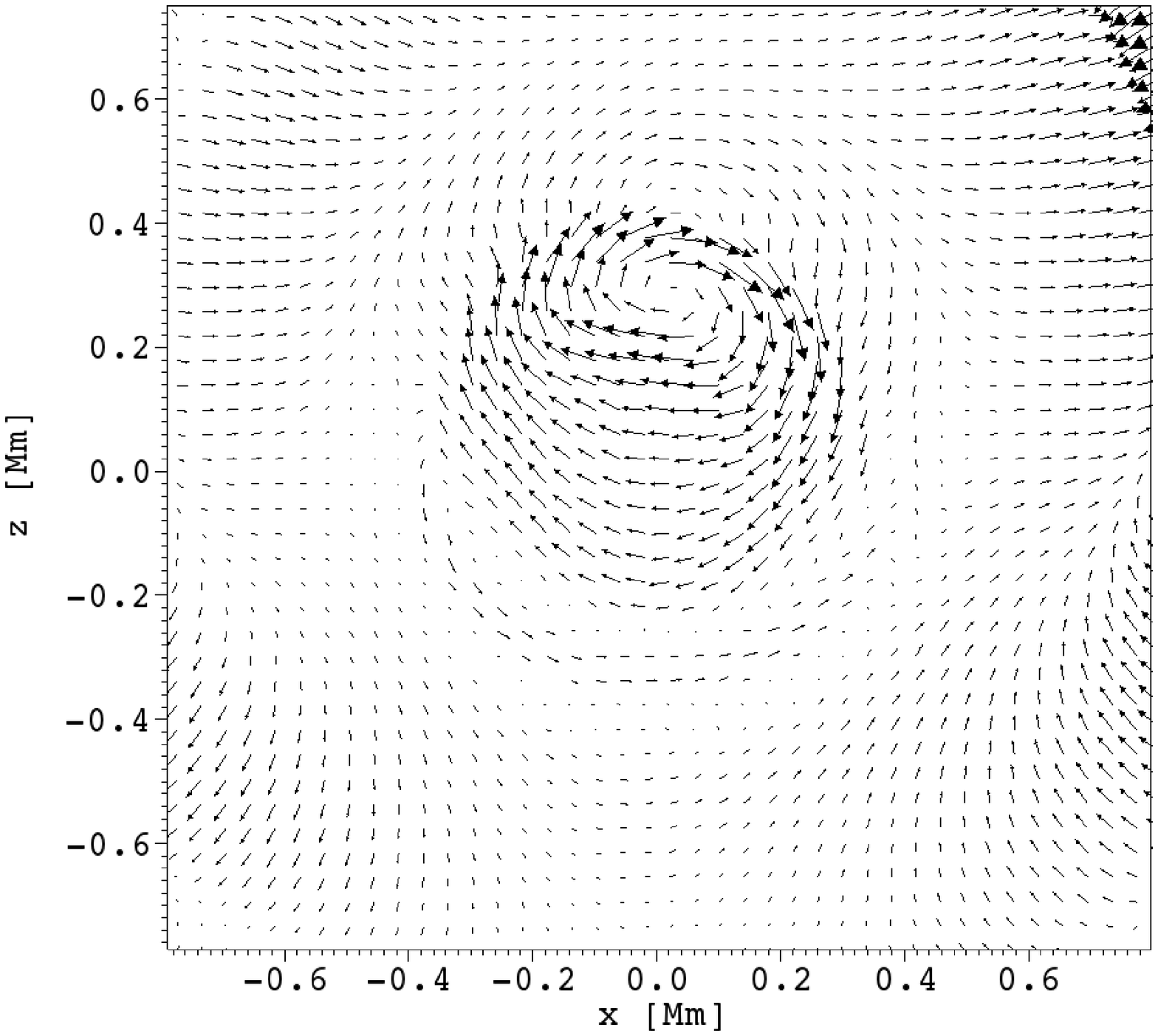}
\includegraphics[width=8.5cm]{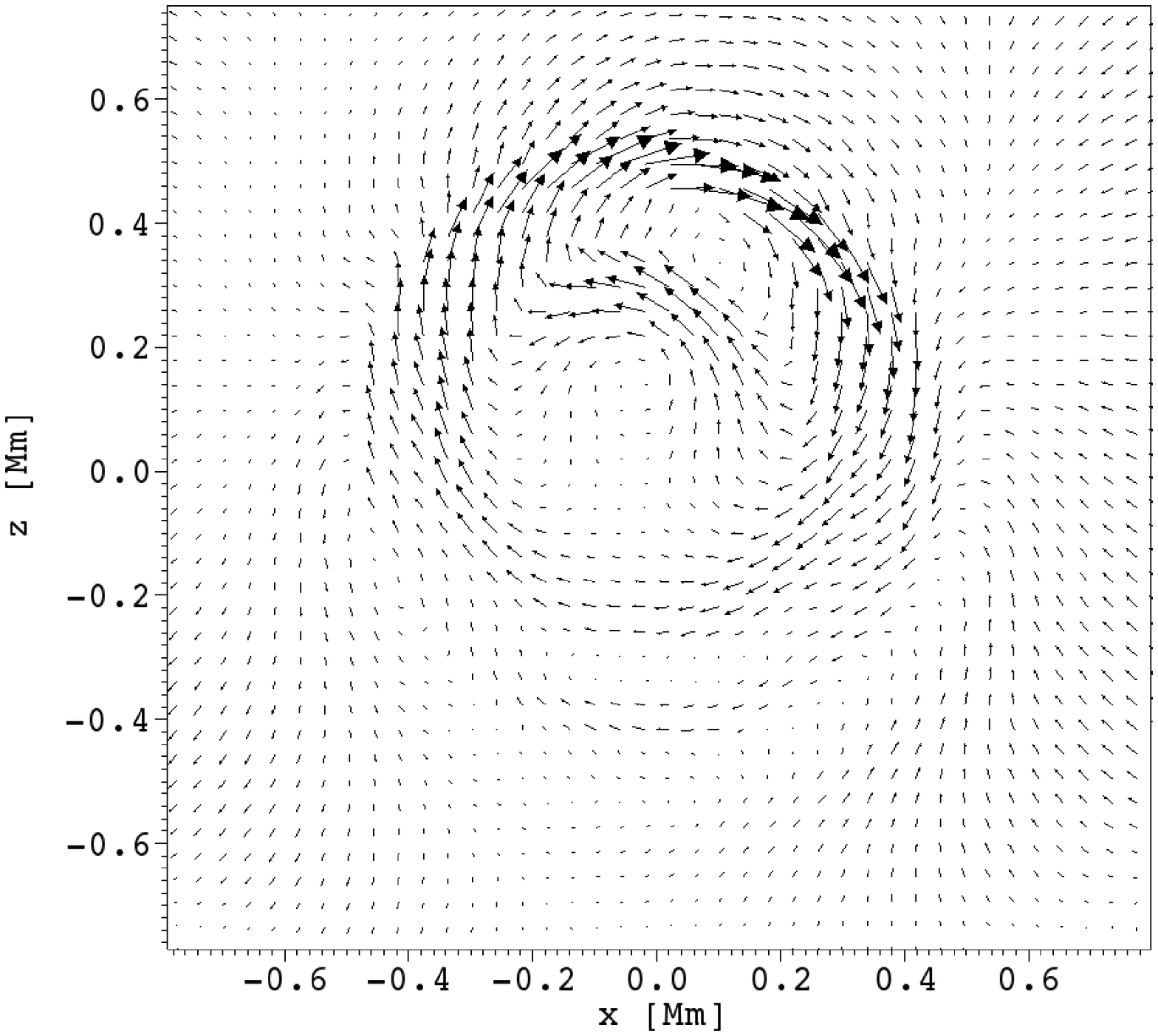}
\includegraphics[width=8.5cm]{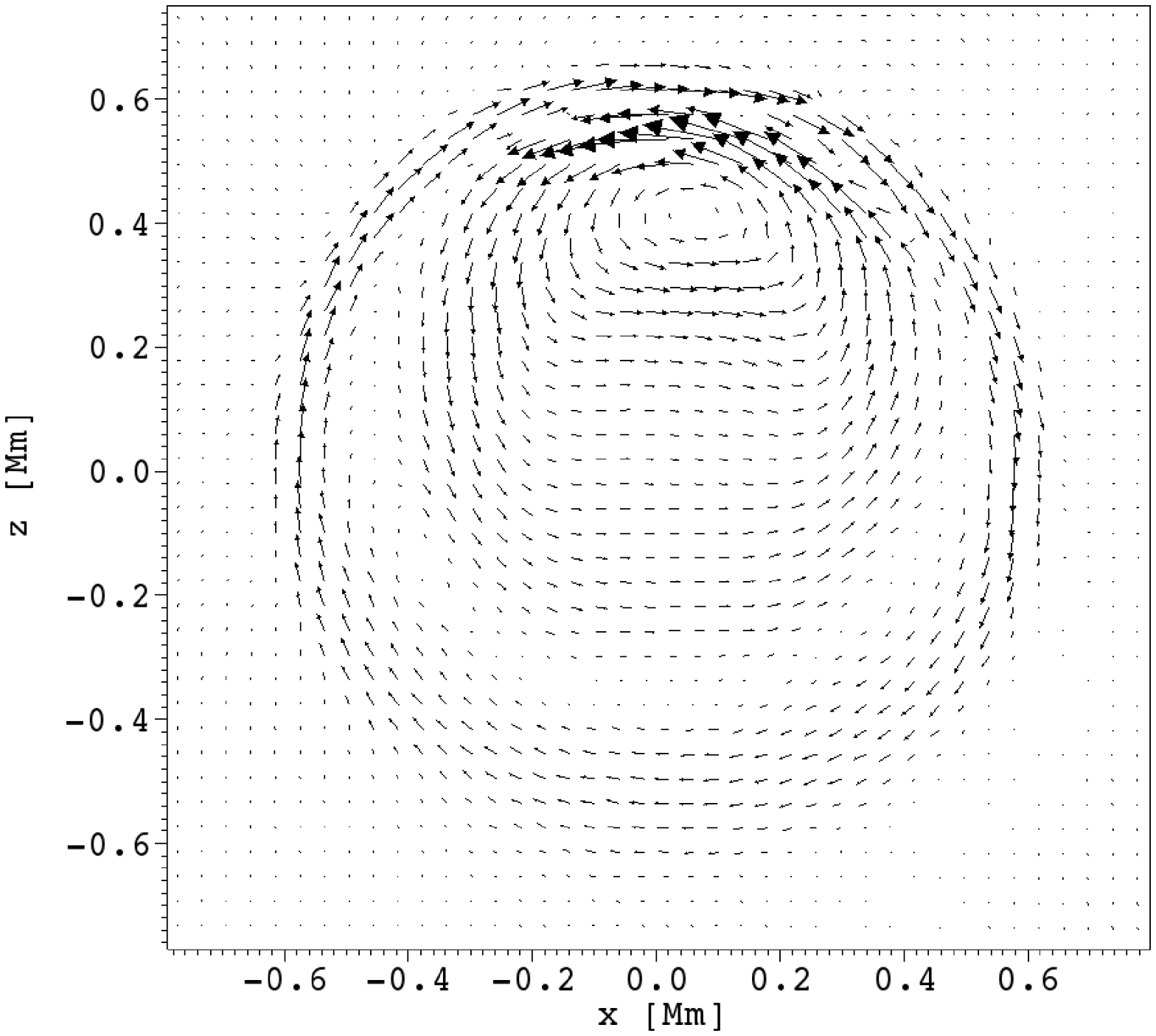} 
\caption{Snapshots of velocity vectors (in arbitrary units) shown at $y=1.5$~Mm at times $t=125$~s (top left), $200$~s (top right), $300$~s (bottom left), and $550$~s (bottom right) for the case of the off-center initial pulse.}
\label{fig:vert-V-off}
\end{figure*}

Figure~\ref{fig:vert-V-off} shows snapshots of the velocity vectors at several times after the off-center initial pulse was launched.
These vectors reveal a single swirl which evolves during the simulation.
At $t=200$~s, the swirl is present around the point of $(x \approx 0, z \approx 0.3)$~Mm (top right panel).
Solitary swirls are also seen at later times (bottom panels) with the presence of outer counter-rotating eddies.
However, in the case of the centrally-launched initial pulse the scenario is more complex as double eddies are observed (compare with Fig.~\ref{fig:vert-V}).
At $t=300$~s, a weak outer swirl centered on $(x \approx 0, z \approx 0)$~Mm is discernible (bottom left).
Additionally, another counter-clockwise plasma rotation is also visible at this time, located at $(x \approx 0, z \approx 0.3)$~Mm, which may be the result of phase-mixing.
At $t=550$~s, the inner swirl rotates clockwise, while the outer swirl has counter clockwise rotation (bottom right panel of Fig.~\ref{fig:vert-V-off}).
At later times both these swirls (i.e. outer and inner swirls) grow in size. In addition, an extra intermediate (i.e. third) swirl develops.
At $t=550$~s, it is clearly visible that the outer swirl shows counter-clockwise motion, while the middle swirl rotates clockwise. The third swirl, which is located around $(x \approx 0, z \approx 0.5)$~Mm, also exhibits clockwise motion.
It should be noted that radius of the flux tube is $0.3$~Mm and the diameters of these swirls are greater than $0.3$~Mm, showing that the swirls can also exist outside the flux tube.
%
\begin{figure*}
\centering
\includegraphics[width=8.5cm]{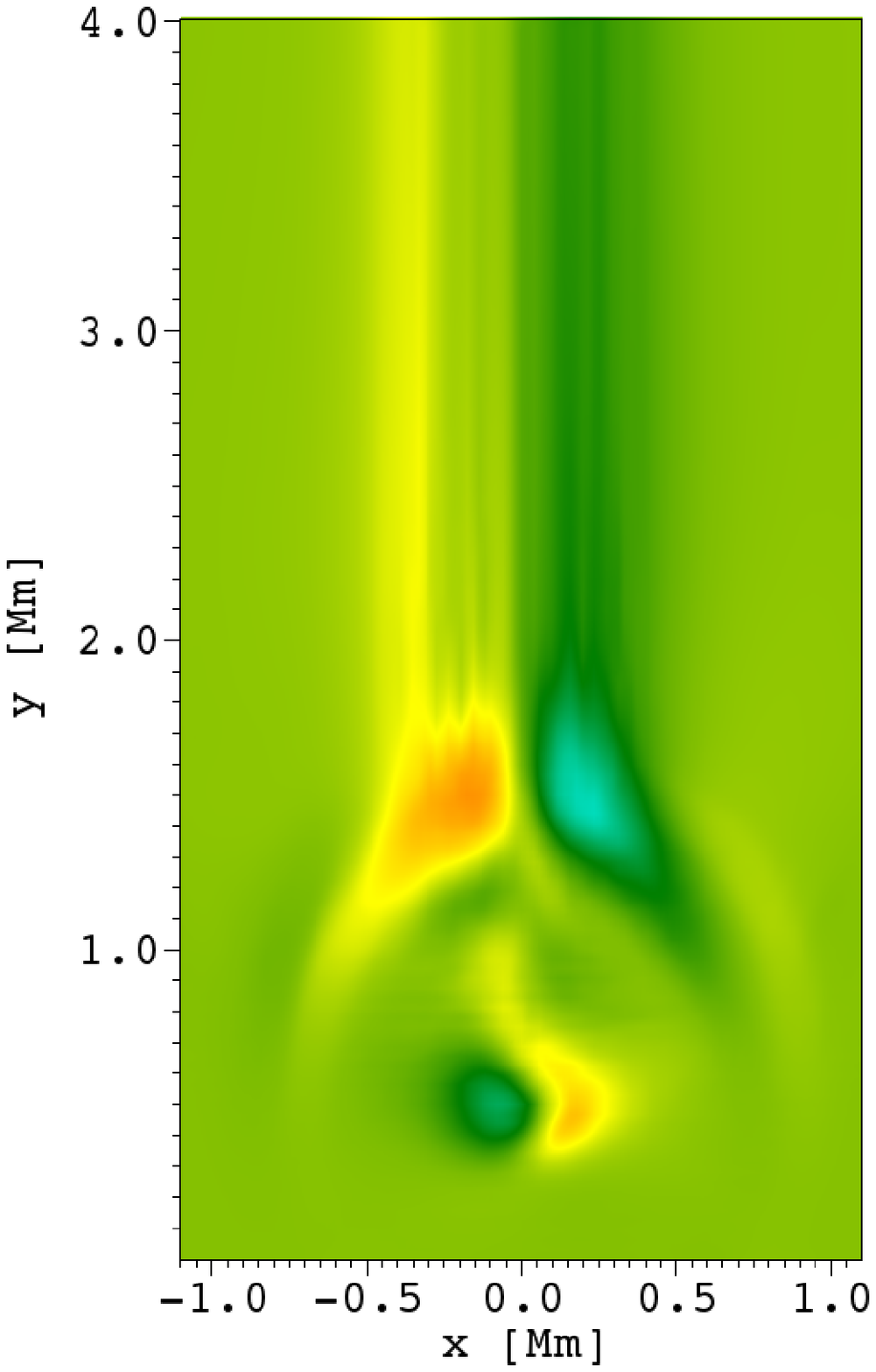}\hspace{-3cm}
\includegraphics[width=8.5cm]{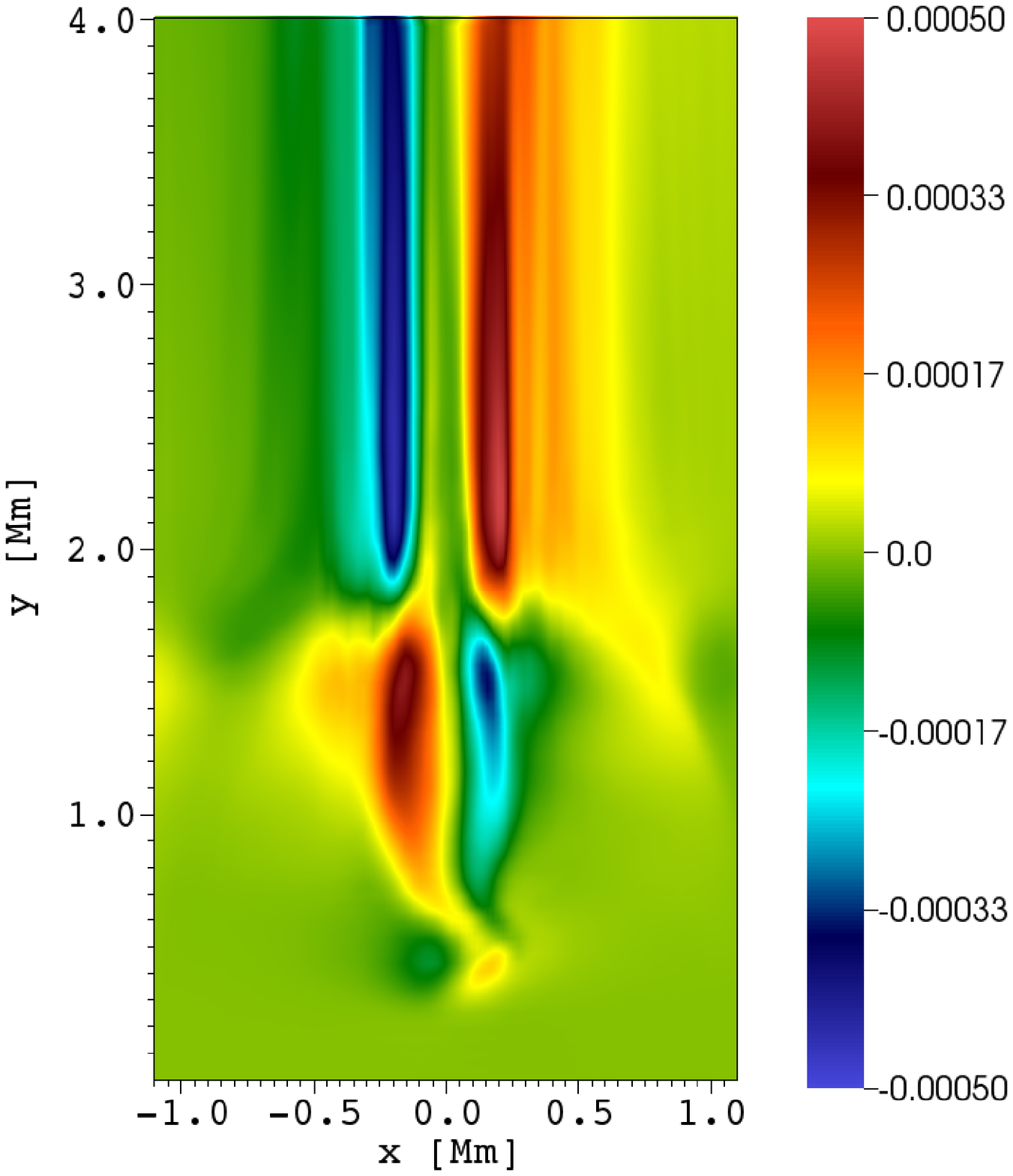}
\includegraphics[width=8.5cm]{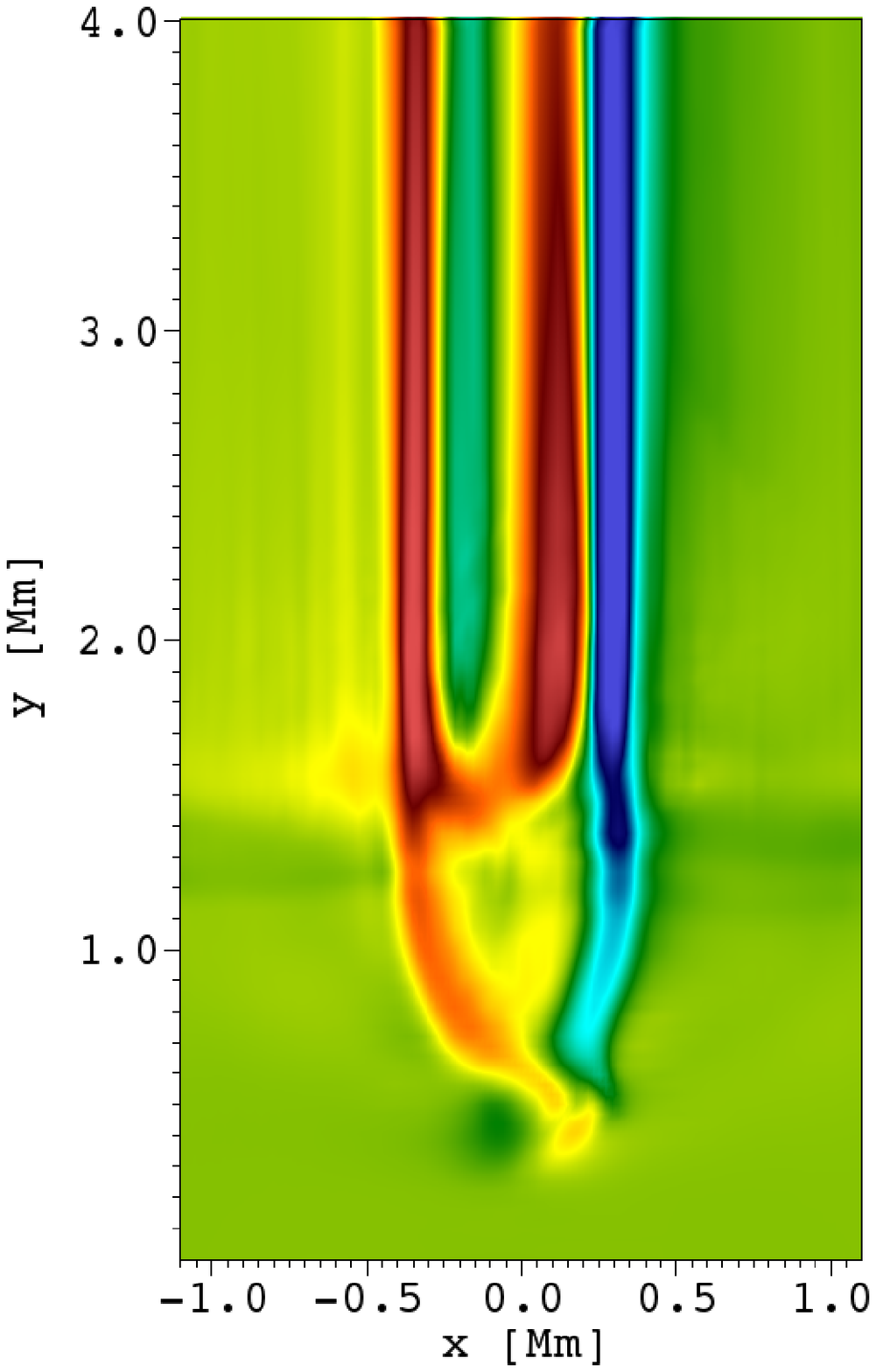}\hspace{-3cm}
\includegraphics[width=8.5cm]{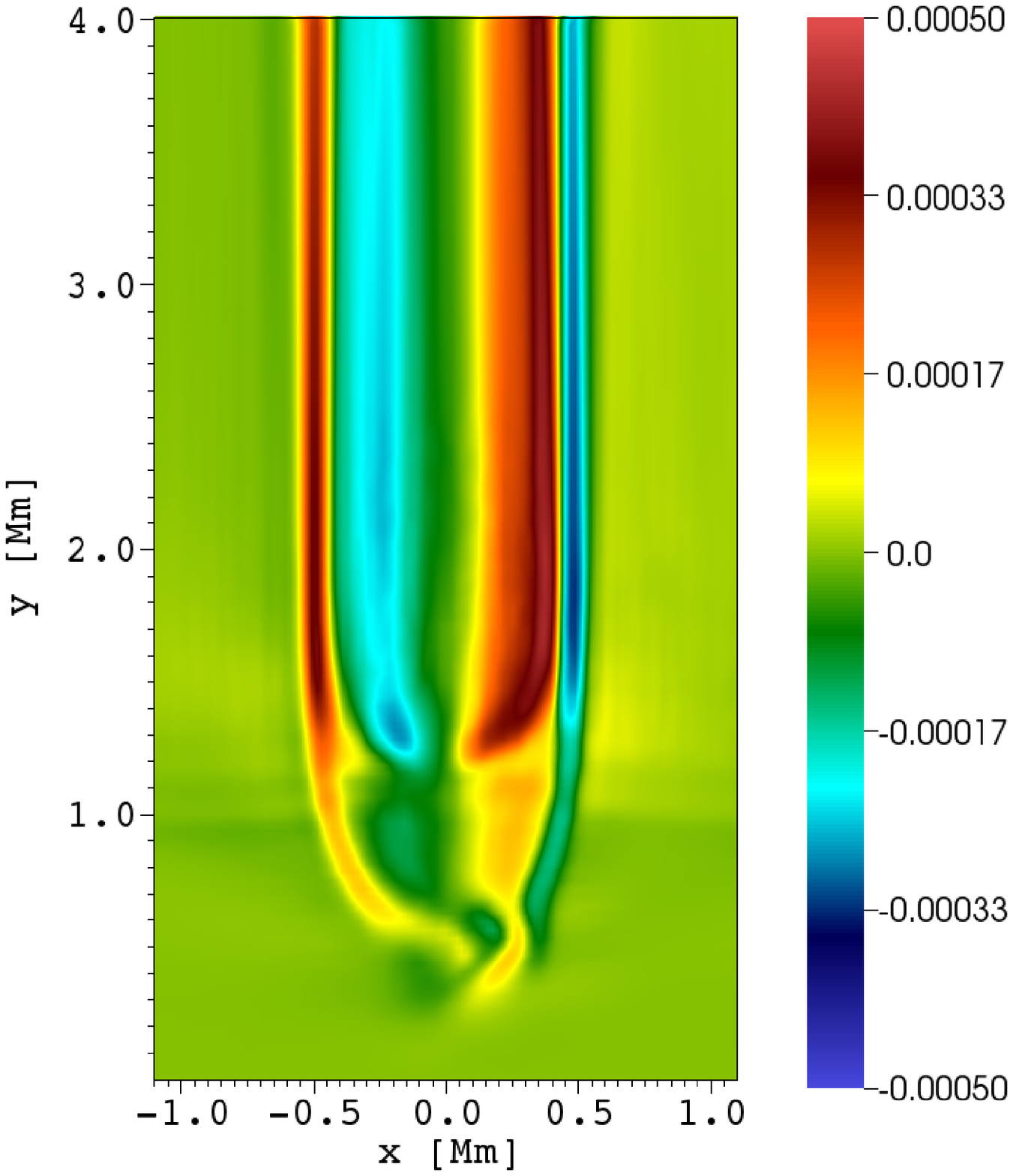}
\vspace{1.5cm}
\caption{Vertical profiles of $V_{\rm z}$ drawn at $z=0.35$~Mm at times $t=125$~s (top left), $200$~s (top right), $300$~s (bottom left), and $550$~s (bottom right) for the case of an off-center initial pulse.
The color map corresponds to the magnitude and direction of velocity, expressed in units of $1$~Mm~s$^{-1}$.}
\label{fig:Vz_z=0.35-off}
\end{figure*}

Figure~\ref{fig:Vz_z=0.35-off} illustrates the temporal variations of vertical profiles of the transversal velocity $V_{z}$, which are shown at $z=0.35$~Mm.
At $t=125$~s, two oppositely directed transversal flows are again visible, a typical signature of rotational motion which indicates the presence of azimuthal Alfv{\'e}n waves as in the case of centrally-launched initial pulse.
At later times, it is seen that the vertical rotating flow expands in the horizontal and vertical directions (bottom panels). 
It is also seen that the alternating fringes of oppositely directed flows are present as in the case of the centrally-launched pulse.
%
\begin{figure*}
\centering
\includegraphics[height=9.5cm]{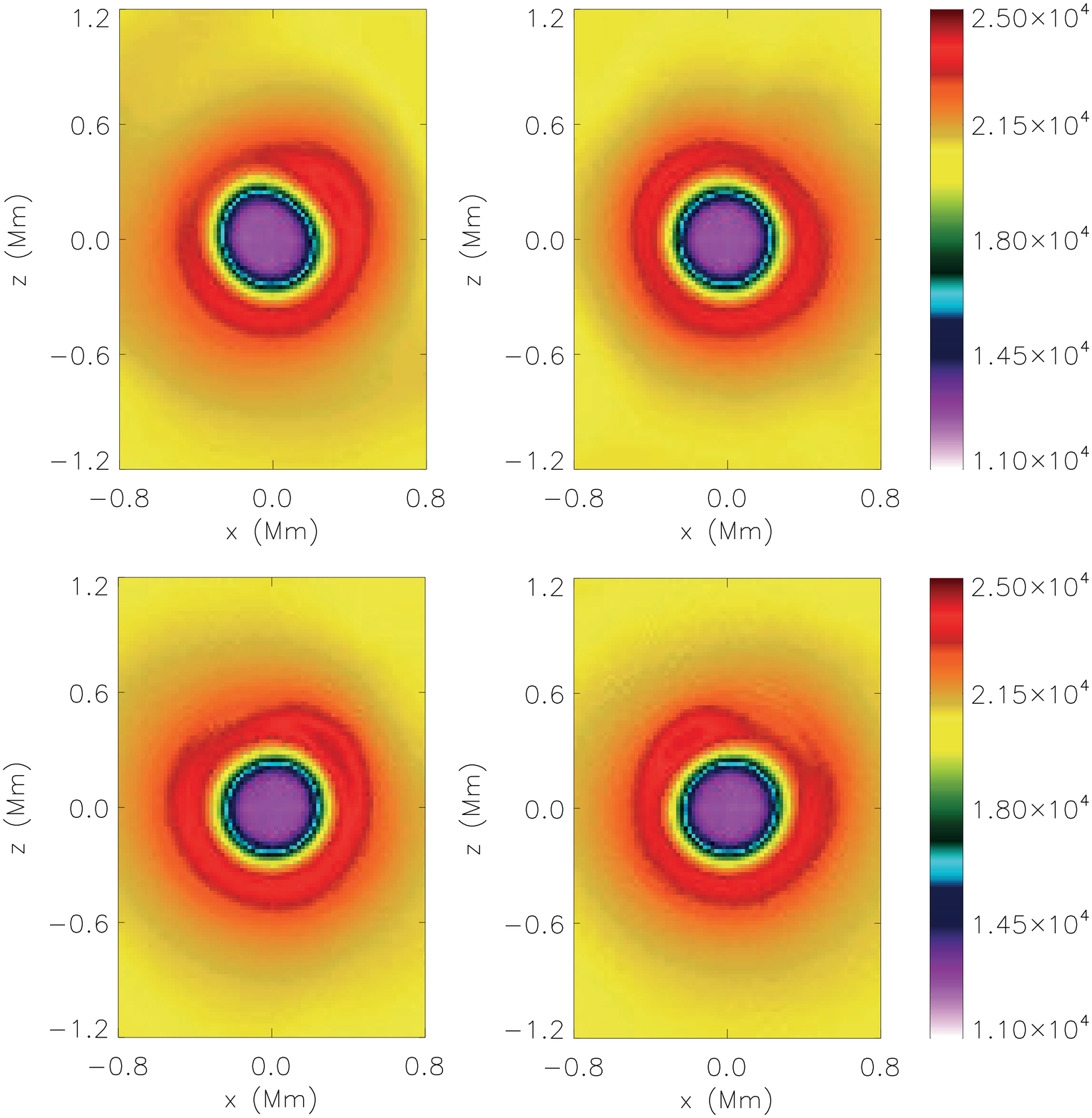}
\caption{Horizontal profiles of mass density $\varrho$ drawn at $y=1.5$~Mm at times $t=300$~s (top left), $400$~s (top right), $500$~s (bottom left), and $600$~s (bottom right) for the case of an off-center initial pulse.
The color map corresponds to the magnitude of $\varrho$ which is expressed in units of $10^{-12}$ kg m$^{-3}$.}
\label{fig:rho-x-y-off}
\end{figure*}

The solitary swirls can also be traced in the horizontal profiles of mass density, $\varrho (x,y=1.5,z)$, shown in Fig.~\ref{fig:rho-x-y-off}.
At $t=300$~s, the profile of $\varrho(x,y=1.5,z)$ reveals inner and outer swirls of ellipsoidal-like shape; the inner (outer) swirl has its longest axis directed to the top left (top right) corner of the panel.
At $t=400$~s, a quasi-circular shape for both outer and inner swirls is seen, while at $t=600$~s the inner oval-shape is directed towards the top right corner of the panel.
The outer oval is oriented along the perpendicular direction. 
Such scenario of quasi-circular shapes corresponds to the $m=0$ Alfv\'en mode 
and it differs from the case of the centrally-launched initial pulse in the kink mode and magnetic shells associated with $m=1$ Alfv{\'e}n mode are well seen (c.f. Fig.~\ref{fig:vert-B}).
So, in the off-central case, we have the dominance of Alfv{\'e}n waves, which transport energy into the solar corona.
\begin{figure*}
\centering
\includegraphics[height=7.1cm]{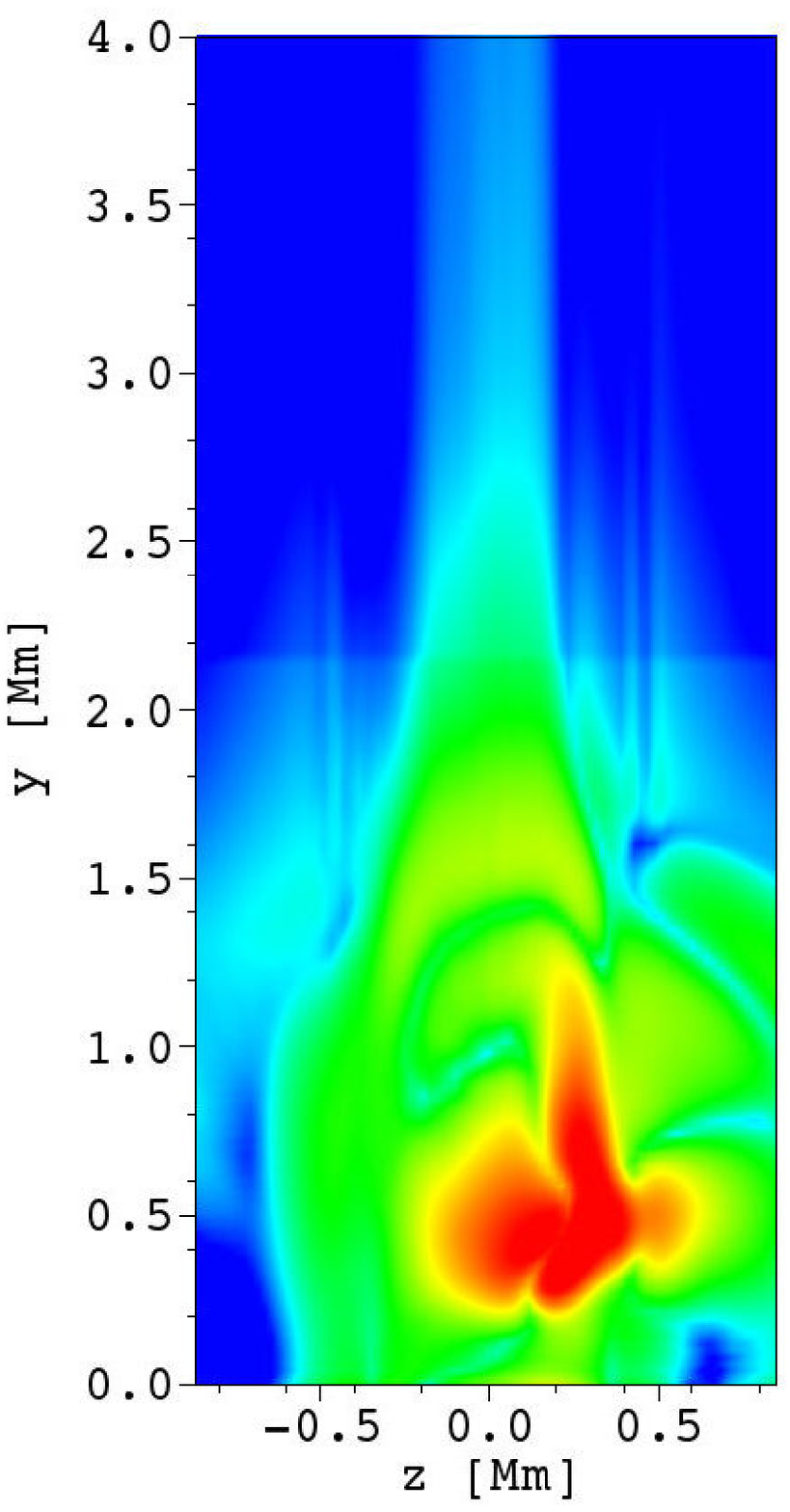}
\includegraphics[height=7.1cm]{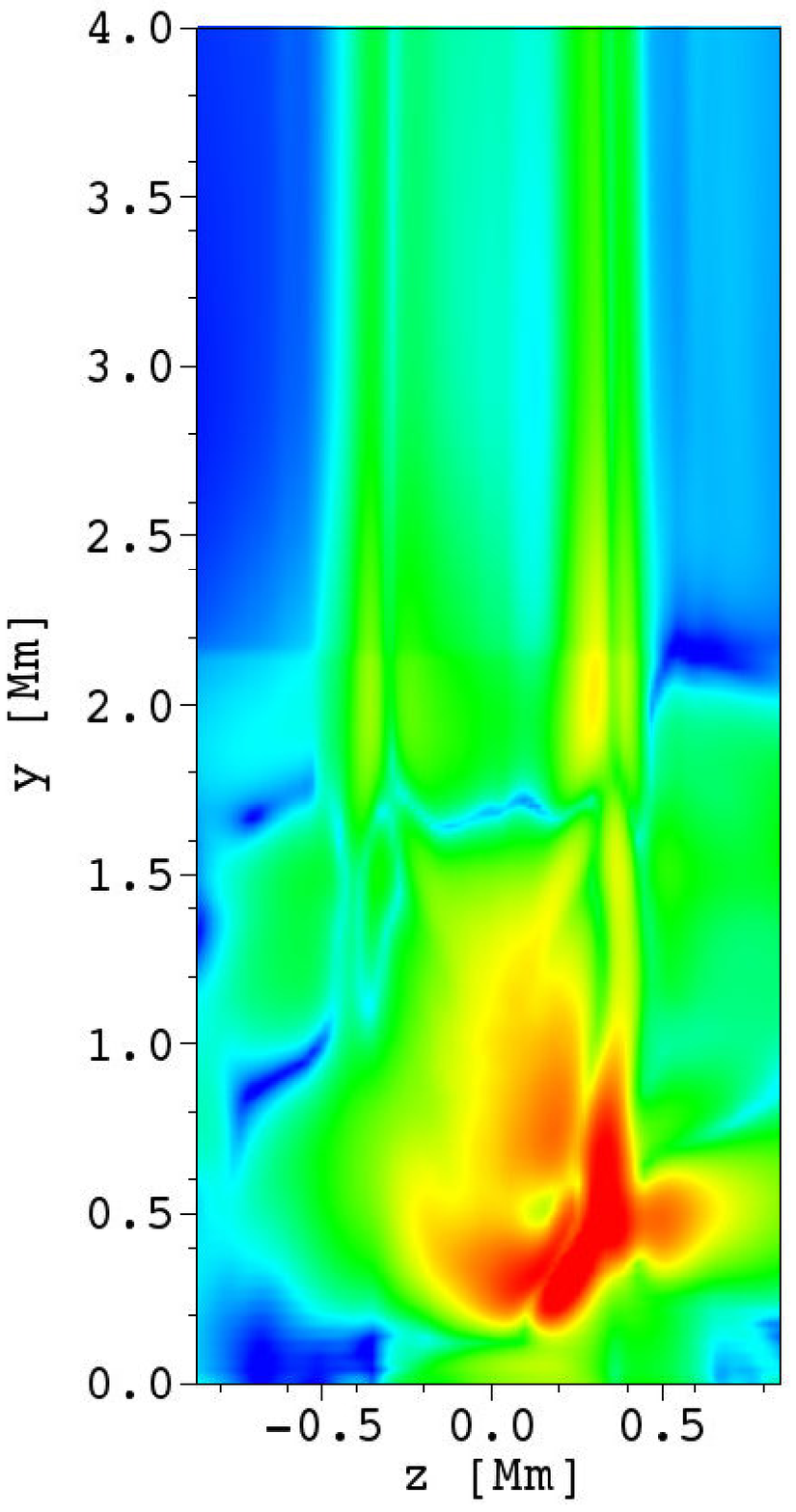}
\includegraphics[height=7.1cm]{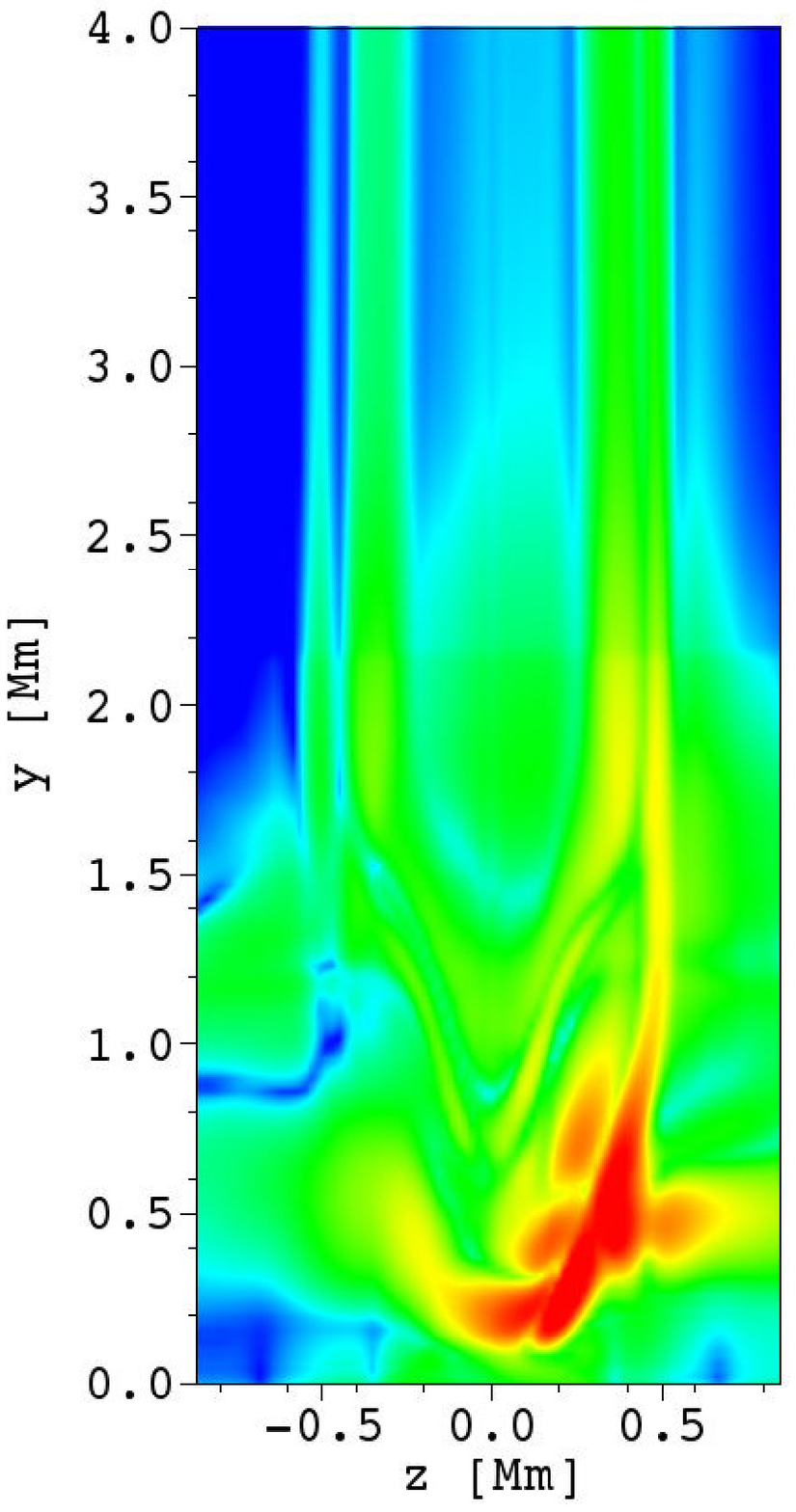}
\includegraphics[height=7.1cm]{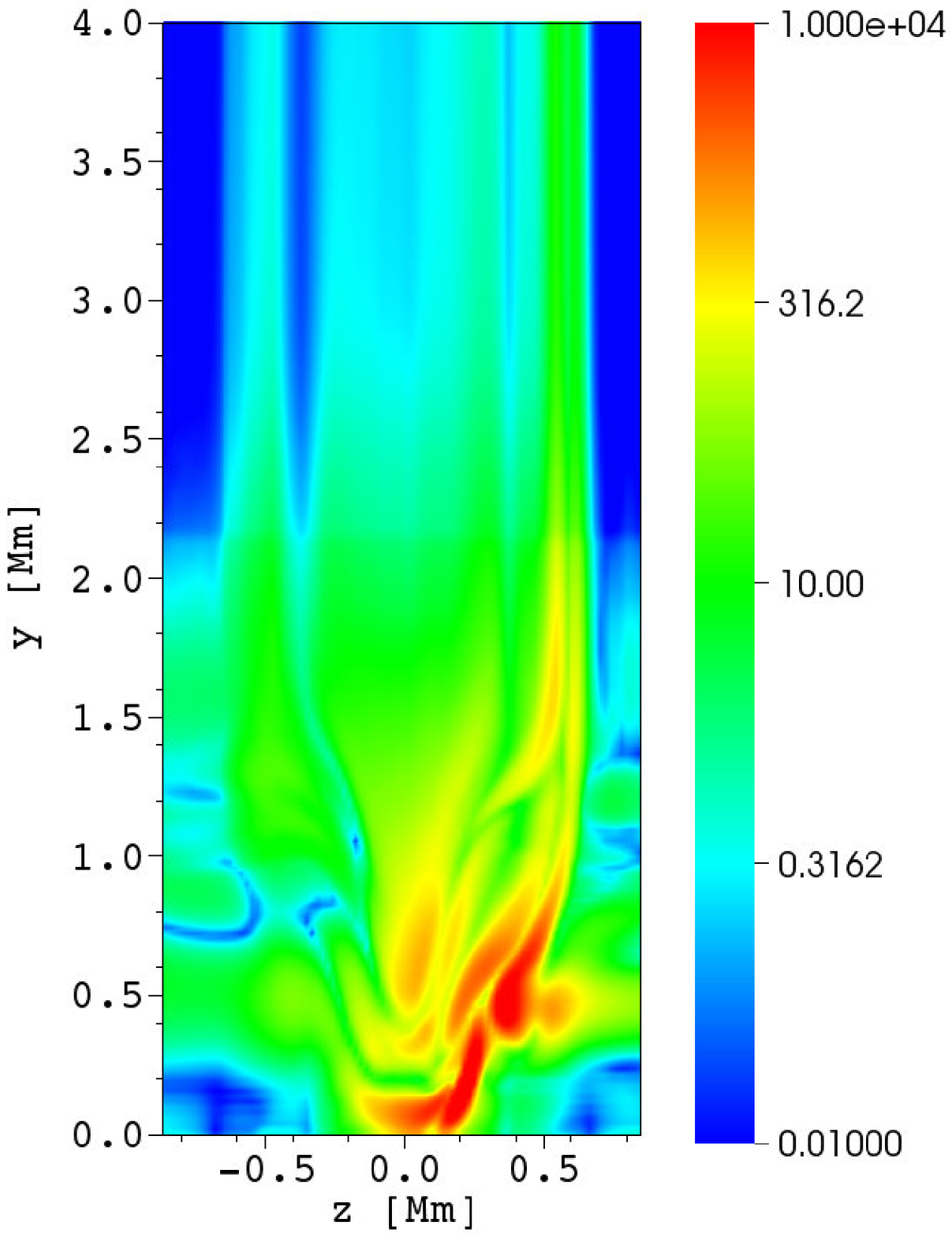}
\caption{Temporal evolution of the energy flux,
$F \approx 0.5 \varrho_{0} c_{\rm A} V^{2}$,
expressed in units of W~m$^{-2}$, drawn at $x=0$~Mm at times (left to right) of $t=125$, $200$, $300$, and $550$~s for the case of an off-center initial pulse.}
\label{fig:fluxoffcentral}
\end{figure*}

The dominance of Alfv{\'e}n wave causes more field-aligned energy transport locally in the flux tube.
It is clearly seen that above the off-central location a sufficient amount of energy flux ($10^{4}$~W~m$^{-2}$) is present in the flux tube at the chromospheric heights to balance its losses (and $\sim 3 \times 10^2$ at $y=2.5$~Mm).
Therefore, the vortices and associated waves are very significant in channelling the energy to the overlying solar atmosphere to heat it locally \citep{Wed2012}.
\section{Summary}\label{sect:summary}
This paper presents the stringent 3D numerical modeling of a solar flux tube with an adaptation of the FLASH code, which in the case of centrally-launched initial pulse in the horizontal component of velocity, demonstrates the generation of two coupled fast magnetic swirls which are associated with coupled kink, $m=1$ or $m=2$ Alfv\'en waves in the solar flux tube.
In the case of the off-centrally-launched initial pulse the $m=0$ Alfv\'en mode and kink waves are triggered.

As a result of the presence of various magnetic structures, the solar photosphere is a very complex dynamic region. 
Horizontal plasma flows are an integral part of the solar photosphere and they are responsible for various kind of dynamics there \citep[e.g.][]{Svanda2006, Zhao2007,Verma2011,Rou2012,Rou2013,Svanda2013}.
Vortex structures are complex phenomena and have attracted significant attention in this high resolution era of observations \citep[e.g.][and references therein]{Wed2009,Wed2012,Wed2013}.
In the context of the formation of these vortex structures, which is crucial to reveal their nature and associated dynamics in the solar atmosphere, 
convective flow is the most promising mechanism to account for their generation \cite[e.g.][]{Brandt1988, Bonet2008,Wed2009,Wed2012,Wed2013}.
It is also reported that these vortex structures can efficiently generate MHD waves \citep[e.g.][]{Erdelyi2007,Jess2009,Fedun2011}. 
However, \citet{Murawski2015} proposed a unified model based on the interaction of a flux tube with a horizontal flow that can produce vortex structures as well as MHD waves.
Here, we have explored this unified model of vortex formation in two different scenarios.
In the case of the centrally-launched initial velocity pulse, the double swirls along with some intermediate complex motions (i.e. formation of four swirls) result from a mere transversal perturbation. 
These double swirls are to some degree similar to the velocity field structure detected by \citet{Steiner2012}. 
Alfv{\'e}n as well as kink waves are also triggered along with the formation of double swirls due to the interaction between the flux tube and horizontal flows.
The Alfv{\'e}n waves reach the solar corona, while the kink waves are attenuated as they propagate.
In the case of coronal kink oscillations, the damping profile can be exploited for seismology to reveal the transverse structure of the flux tube \citep{2013A&A...551A..40P,2016A&A...589A.136P,2017A&A...600A..78P}.
However, our chromospheric flux tube is more complicated, particularly since it acts as an anti-waveguide and so also leaks energy into the ambient medium.
The energy flux carried by these waves is sufficient to heat the solar corona.
In the case of the centrally-launched initial pulse, the developed model of two coupled fast swirls in the 3D flux tube with upper atmospheric responses can be considered to be similar to the dynamics of dipole-shaped chromospheric swirls.
Our model, therefore, can be considered as a manifestation of dipole-like swirling motions in the solar atmosphere \citep{Wed2013}.
However, in the case of an off-central pulse, the formation of a major single swirl is visible and mainly Alfv{\'e}n waves are present in the system. 
Interestingly, only these waves carry a sufficient amount of energy to heat the solar corona.

In conclusion, our 3D numerical simulations contribute to understanding the physical conditions in the solar atmosphere by using more realistic temperature and related conditions within a flux tube rooted at the photosphere and fanning out up to inner corona.
Our model mimics the formation of confined vortices and MHD waves, and their responses into the chromosphere, transition region, and inner corona. 
The centrally-launched velocity pulse leads to the formation of double swirls with the natural occurrence of Alfv{\'e}n and kink waves, while the off-central pulse results in the formation of a major single swirl 
with the natural occurrence of predominantly Alfv{\'e}n waves.
Kink waves leak out of the flux tube which constitutes 
an anti-waveguide for fast magnetoacoustic waves.
Therefore, it suggests that the change in the interaction region between the flux tube and horizontal flow can significantly alter this dynamics. 
It should be noted that a very high resolution is needed to observe such fine dynamics in the form of these vortices.
%
%
\section*{Acknowledgeents}
The authors express their thanks to the anonymous referee and Drs. Oskar Steiner and Gary Verth for their comments on earlier version of this draft. 
This work was supported by the project from the National Science Centre, Poland, (NCN) Grant no. 2014/15/B/ST9/00106. 
PJ acknowledges the support from grant 16-13277S of the Grant Agency of the Czech Republic.
AKS acknowledges the ISRO-RESPOND and SERB-DST projects.
DJP is supported by the European Research Council under the \textit{SeismoSun} Research Project No. 321141. 
VF would like to thank the STFC for financial support received. 
The software used in this work was in part developed by the DOE-supported ASCI/Alliance Center for Astrophysical Thermonuclear Flashes at the University of Chicago. 
Numerical simulations were performed on the LUNAR cluster at Institute of Mathematics of University of M. Curie-Sk{\l}odowska, Lublin, Poland. 
The numerical data was visualized with the use of IDL, ViSiT, and Vapor graphical packages.
%
%
%

%
%
\appendix
\section{Equilibrium plasma pressure and mass density}
The equilibrium plasma pressure and mass density are described by the following formulas 
(Solov'ev 2010; Solov'ev \& Kirichek 2015; Murawski et al. 2015a): 
\beqa
\label{eq:33pe}
p(r,y) = p_{\rm h}(y) - \frac{1}{\mu}\left[ \frac{1}{2r^2}\left( \frac{\partial\Psi}{\partial r} \right)^2 + \int_\infty^r\frac{\partial^2\Psi}{\partial y^2}\frac{\partial\Psi}{\partial r}\frac{dr}{r^2} \right]\, ,	\\
\label{eq:33rhorho}
\varrho(r,y)  =  \varrho_{\rm h}(y)\,\nonumber \\
+  \frac{1}{\mu g}\frac{\partial}{\partial y}\left[ \frac{1}{2r^2}\left(\left( \frac{\partial\Psi}{\partial r} \right)^2 - \left( \frac{\partial\Psi}{\partial y} \right)^2 \right)
+  \int_\infty^r\frac{\partial^2\Psi}{\partial y^2}\frac{\partial\Psi}{\partial r}\frac{d r}{r^2} \right]  \\
 -  \frac{1}{\mu g r}\frac{\partial\Psi}{\partial y}\frac{\partial}{\partial r} \left( \frac{1}{r}\frac{\partial\Psi}{\partial r} \right)\, ,\nonumber
\eeqa
where
\beq
\Psi(r,y) = \int_0^r B_{\rm y} r' dr'\, ,
\label{eq:33Psi}
\eeq
is the normalized (by $2\pi$) magnetic flux function, and 
hydrostatic gas pressure $p_{\rm h}$ and mass density $\varrho_{\rm h}$ are given by 
\beq
p_{\rm h}(y) = p_{\rm 0} \exp\left[ -\int_{y_{\rm r}}^{y} \frac{dy'}{\Lambda(y')} \right]\, , \hspace{3mm}	
\varrho_{\rm h}(y) = \frac{p_{\rm h}(y)}{g\Lambda(y)}
\eeq
%
with 
\begin{equation}
\Lambda(y) = \frac{ k_{\rm B}T_{\rm h}(y) }{ mg }
\end{equation}
being the pressure scale-height that we specify by adopting the semi-empirical model of plasma temperature brought by Avrett and~Loeser (2008).  


\bsp	
\label{lastpage}
\end{document}